\newcommand{\hi}{{\sc H\,i}\,}
\newcommand{\Mhi}{M$_{\text{\sc H\,i}}$\,}
\documentclass{aa}  

\usepackage{color}
\usepackage{graphicx}
\usepackage{subfig}
\usepackage{multirow}
\usepackage{caption}
\usepackage{float}
\usepackage{txfonts}
\usepackage{hyperref}
\usepackage{lscape}
\begin{document}

\title{The Arecibo Galaxy Environment Survey (AGES) XI: \\
the expanded Abell 1367 field}

\subtitle{Data catalogue\thanks{Tables with properties of \hi detections and their optical counterparts are available in electronic form
at the CDS via anonymous ftp to cdsarc.u-strasbg.fr (130.79.128.5)
or via http://cdsweb.u-strasbg.fr/cgi-bin/qcat?J/A+A/ \\ Throughout this article, clicking on the GitHub logo will open the Python script used for executing the calculation or producing the figure being discussed.} and \hi census over the surveyed volume}

   \author{Boris Deshev\inst{1}
          \and Rhys Taylor\inst{1}%\fnmsep
          \and Robert Minchin\inst{2}
          \and Tom C. Scott\inst{3}
		  \and Elias Brinks\inst{4}
          }

   \institute{Astronomical Institute, Czech Academy of Sciences, Bo\u{c}n\'i II 1401, CZ-14131 Prague, Czech Republic\\
              \email{deshev@asu.cas.cz}
         \and
         		Stratospheric Observatory for Infrared Astronomy/USRA, NASA Ames Research Center, MS 232-12, Moffett Field, CA 94035, USA
		 \and
				Institute of Astrophysics and Space Sciences (IA), Rua das Estrelas, P-4150-762 Porto, Portugal
		 \and
		 		Centre for Astrophysics Research, University of Hertfordshire, College Lane, Hatfield AL10 9AB, UK
                        }         

   \date{}

% \abstract{}{}{}{}{} 
% 5 {} token are mandatory
 
  \abstract
   {Many galaxy properties are known to correlate with the environment in which the galaxies are embedded. Their cold, neutral gas content, which is usually assessed through 21 cm \hi observations, is related to many other galaxy properties as it is the underlying fuel for star formation. With its high sensitivity and broad sky coverage the blind Arecibo Galaxy Environment Survey (AGES) brings significant improvement to the census of \hi properties of galaxies in a wide range of environments, from voids to the core of a massive cluster. Here we present an \hi census over a volume of $\sim$ 44000 Mpc$^{3}$ towards the merging cluster Abell 1367 and extending well beyond.}
   {To measure the effects that different environments have on the \hi content of their constituent galaxies.}
   {We use AGES- a deep, blind, \hi survey carried out with the Arecibo radio telescope, which covers 20 square degrees on the sky centred on the merging cluster Abell 1367, mapping the large-scale structure (LSS) surrounding the cluster out to $cz$ = 20000 km s$^{-1}$. The survey is sensitive down to a column density of N$_{\text{\sc H\,i}}$ = 1.5 $\times$ 10$^{17}$ cm$^{-2}$ for emission filling the beam and a line width of 10 km s$^{-1}$. As an approximate mass sensitivity limit, a member of A1367 (at a distance of 92 Mpc), containing M$_{\text{\sc H\,i}}$ = 2.7 $\times$ 10$^{8}$ M$_{\odot}$ distributed over a top-hat profile of 50 km s$^{-1}$ width would be detected at 4$\sigma$. The results are analysed in combination with optical spectroscopy data from SDSS which we use to estimate the local galaxy density based on the Voronoi-Delaunay method.}
   {We present the results of the complete AGES survey of the A1367 field. In total, we detect 457 \hi sources, 213 of which are detected for the first time by the AGES survey, and 134 of which are presented in this article for the first time. Of the 457 detections, 225 are in the cluster and 232 are in the remaining volume surveyed. Here we present the full catalogue of \hi detections and their basic properties, including optical ones. We concentrate on the difference between the cluster and the foreground and background LSS, revealing a continuous correlation of \hi-detected fraction (and \hi deficiency) with local galaxy density, independent of  global environment.}
   {}


   \maketitle
%-------------------------------------------------------------------

\section{Introduction}
In recent decades, the availability of data from blind, all-sky surveys like the Sloan Digital Sky Survey \citep[SDSS, ][]{SDSS1} and The Arecibo Legacy Fast ALFA Survey \citep[ALFALFA, ][]{alfalfa1} have enabled an ever growing appreciation of the effects of environment on the evolution of galaxies \citep{Boselli&Gavazzi2006, BoselliGavazzi2014, Boselli2021}. The presence of ram pressure \citep[RPS hereafter,][]{Gunn&Gott1972}, which actively removes gas from galaxies falling onto large clusters, has been unequivocally proven \citep{KenneyvanGorkom&Vollmer2004} and the fine details of its workings have now been observed \citep{Arrigoni2012, Boselli2018, Jachym2019}. This is now known to be an important factor in explaining the gas content and star formation properties of cluster galaxies \citep{Cortese2021}.

On the other hand, the vast majority of galaxies do not reside in clusters, which have halo masses of log M$_{h} \geq$ 14 M$_{\odot}$, where RPS could realistically have detectable effects on a large galaxy's \hi disc. Taking all galaxies with $z \leq$ 0.1 from the the Max Planck Institute for Astrophysics - Johns Hopkins University (MPA-JHU) value-added galaxy catalogue \footnote{https://www.sdss.org/dr12/spectro/galaxy\_mpajhu/} based on SDSS shows that $\sim$ 40\% of them show little-to-no ongoing star formation. If we combine the same data set with the cluster and group catalogue of \citet{Tempel2017}, we see that clusters with M$_{h} \geq$ 14 host only around 8\% of all galaxies. Understanding what determines the properties of non-cluster galaxies has proven much more difficult \citep{Cortese2021}.

For a long time it has been known that the closed box model for galaxy evolution does not match the observed metallicity of Milky way stars \citep{Schmidt1963, Tinsley1980}, which suggests a continuous accretion of low-metallicity gas from the circumgalactic medium. This is supported by the gas depletion timescales, which for the $z$ = 0 galaxy population are significantly shorter than the Hubble time \citep{Saintonge2017}. Gas accretion must therefore be the dominant environmental effect that impacts all galaxies. The absence of accretion is termed starvation or strangulation and has been suggested as an environmental effect by \citet{Larson1980} and confirmed to be affecting most passive galaxies in the local Universe \citep{Peng2015}. This prevention of gas accretion leads to a slow decline in star formation rate until its complete cessation $\sim$ 4 Gyr later. 

Studies of the effects of environment use several different definitions of environment that are sensitive to the working of different physical processes. On one side is the global environment, which is usually the membership of the galaxies to a cluster or a group with a subdivision between central and satellite galaxies within those units. There are a number of catalogues of groups, clusters, and superclusters based on SDSS \citep[e.g.][]{Yang2007, Tago2010, Liivamagi2012, Tempel2017}. The local environment is usually assessed by some measure of the local galaxy number density, be it projected on the sky or in 3D.

The cluster A1367 is an ongoing merger of multiple structures \citep{Bechtold1983, Donnelly1998, Girardi98, Sakai2002, Gavazzi2003b, Cortese2004}. Its properties have been extensively studied and have revealed the presence of a plethora of physical processes related to cluster infall, such as merger shocks \citep{Ge2019}, X-ray tails \citep{Ge2021}, multiple jelly-fish galaxies \citep{Scott2010, Scott2012, roberts2021}, merger induced star formation \citep{Gavazzi2003b}, ram pressure stripping and tidal interactions \citep{Gavazzi2001, Scott2018}, and preprocessing \citep{Cortese2006}. The cluster is also part of the 'Great wall' filament that connects it to the Coma cluster \citep{Zabludoff1993}.

For consistency with previous AGES publications, throughout this article we use standard $\Lambda$CDM with H$_{0}$ = 71 km s$^{-1}$Mpc$^{-1}$. Thus, 1\arcsec = 0.44 kpc at the distance of A1367 (92.8 Mpc), and the full width at half maximum (FWHM) of the resampled Arecibo beam ($3.4\arcmin$) at the same distance is $\sim$ 90 kpc.

\section{Observations and data reduction}
The AGES used the seven-beam Arecibo L-Band Feed Array (ALFA) instrument which has a hexagonal footprint on the sky. The FWHM of the central beam is $\sim$ 3.3 $\times$ 3.8\arcmin. The observing strategy is described in detail in \citet{auld2006}, \citet{cortese2008}, and \citet{davies2011}. Briefly, AGES is a drift-scan survey, with the telescope driven to the start position and the sky allowed to drift overhead for the duration of the scan. Each scan therefore covers a swathe of right ascension (R.A.), with subsequent scans staggered by approximately one-third of a beam width ($\sim$ 1\arcmin) in declination (Dec) to form a Nyquist-sampled map. A source above Arecibo takes 12 s to cross the beam. With seven beams in the ALFA instrument, each point is scanned 25 times, giving an on-source integration time of 300 s. Each beam integrates data from two polarisations, once per second, over 4096 channels, with a spectral resolution equivalent to 5.5 km s$^{-1}$ per channel which is three-point Hanning smoothed to 10 km s$^{-1}$ for greater sensitivity. The bandwidth spans the velocity range -2000 to +20000 km/s using the WAPP spectrometers. The output cube is the average of the two polarisations.

Observations for the A1367 field were taken in April-May 2006, April 2007, November 2011, January-February 2012, May-June 2012, April-May 2013, December 2013, February-March 2015, May-June 2015, December 2016 ? February 2017, June 2017, January-June 2018, and December 2018 ? February 2019. Observations were prioritised to begin with a central 5 $\times$ 1 degree strip centred on the main body of the Abell 1367 cluster. These initial results were fully described in \citet{cortese2008}. 

Here we present the full 5 $\times$ 4 degree field, spanning R.A. (11:33:32 ? 11:54:39) and Dec (17:45:00 ? 21:54:00). Each drift scan covered the full 5 degrees of R.A. (20 minutes of time). The data were reduced with \textsc{livedata} and \textsc{gridzilla} (described in \citet{barnes2001}). \textsc{Livedata} is used to measure and subtract the bandpass and \textsc{gridzilla} combines the scans to produce the final data cube with a circular Gaussian beam with FWHM of 3.4\arcmin. Following \citet{taylor2014}, we found it to be beneficial to further process the data in Python, typically applying a second-order polynomial to the full spectral baseline. This greatly improves the baseline fit (especially with regards to the effects of strong continuum sources) and, as the baseline is large and the polynomial is of low order, it does not introduce any artefacts that could be potential false-positive detections. 

We also apply a MEDMED-style estimator to the spatial bandpass. As described in \citet{putman2002}, this measures the spatial bandpass (the scan in R.A.) in five equal-sized boxes, and subtracts the median of the medians. Although the median of the entire scan is usually an adequate measurement of the bandpass, particularly bright, extended sources can exaggerate the median value above the true value of the bandpass; therefore, using the median of the medians gives a more robust value (see \citet{minchin2010} for an example; further discussion of the combined effects of this processing is given in \citet{taylor2020}).

From measurements of individual sources (see following section), the typical rms of individual Hanning smoothed spectra is $\sim$ 0.6 - 0.7 mJy, equivalent to a 1$\sigma$ column density sensitivity of N$_{\text{\sc H\,i}}$ = 1.5 $\times$ 10$^{17}$ cm$^{-2}$ for emission filling the beam and with 10 km s$^{-1}$ line width \citep{keenan2016}. As an approximate mass sensitivity limit, at the 92.8 Mpc distance of the cluster (following \citet{cortese2008} for consistency with the previous AGES analysis), a 4$\sigma$, 50 km s$^{-1}$ velocity width source with a top-hat profile would have M$_{\text{\sc H\,i}}$ = 2.7 $\times$ 10$^{8}$ M$_{\odot}$.

\subsection{Source extraction}\label{section_source_extraction}
	In this work, we have adopted a source-extraction strategy similar to all previous AGES papers. We catalogue detections using a combination of visual (by eye) and automatic techniques. The visual methods use the visualisation tool FRELLED \citep{taylor2015}. This is a two-stage process. First, FRELLED renders the data as a three-dimensional volumetric cube, allowing the user to interactively create masks to identify and measure potential sources. The advantage of a 3D rendering is that the user can instantly see the full extent of a source in both the spatial and spectral axes, making the masks easy to define. The disadvantage is that the volumetric rendering requires substantial clipping in order to minimise the visibility of the noise, making it unsuitable for finding the faintest sources. We therefore also use a second stage, whereby, after masking the brightest sources in 3D, we scan the cube as a series of 2D slices. This visualisation method does not require such heavy clipping but makes it more difficult (i.e. slower) to define the source masks. The combination of 3D and 2D inspection is therefore designed to optimise the speed and completeness for both bright and faint sources (further details are given in \citet{taylor2015}). The whole identification process was performed independently by two of the authors of the present article (BD and RT). Once complete, FRELLED uses the masks to generate input parameter files for the Miriad task \textsc{mbspect} \citet{sault1995}, which we used to measure the spectral parameters of each source. These input parameters are included in the \hi spectrum files available through the AGES online database\footnote{\url{http://www.naic.edu/~ages/public/}}.\\	

All source-extraction methods suffer from limited completeness and reliability. We supplement our visual inspection with the automatic program GLADoS, which is described in \citet{taylor2012}. In brief, GLADoS searches each spectrum for pixels above a specified S/N threshold (4$\sigma$) which are contained within spectral features above a velocity width threshold (25 km s$^{-1}$). When a candidate is found in the standard averaged cube, GLADoS repeats the search in the same spectrum within each polarisation. As the noise level is slightly higher in the separate polarisations, the S/N threshold is slightly softened to 3.5$\sigma$ and additional Hanning smoothing is applied to each individual cube. \\

We re-wrote the original GLADoS code in Python and added substantial new functionality (to be described in detail in a future paper). One challenge for automatic algorithms, as described quantitatively by \citet{taylor2013}, is their low reliability; it is no advantage that the process is automatic if the initial candidate source list numbers are in the tens of thousands. GLADoS mitigates this through the use of multiple polarisations, and in the new version, by a GUI which allows the user to set the candidate status (accept, reject, or uncertain) interactively. This allows a much more rapid inspection and re-inspection of the automated catalogue. In addition, as with FRELLED, it also outputs an optical image of the SDSS field at the \hi coordinates alongside the \hi spectrum. Although we would prefer the \hi identification to be entirely independent of the optical information, in practise the presence of an optical galaxy at the position of an \hi candidate source (even a weak one) has been found to significantly improve the probability that the candidate is real. Finally, we further reduce the burden of the automatic extractor by first applying the masks generated by the visual search, so that GLADoS avoids repeating the detections already found visually.\\

All sources extracted with both methods are measured using \textsc{mbspect}, assuming they are unresolved. A complete list of the 457 \hi detections and their measured properties is given in the online tables. We assign a descriptive flag to all sources in the catalogue ranging from 0 to 4. These are based on visual inspection performed individually by each member of the collaboration. Inspected were the noise pattern in the data cube around each detection, as well as its spectrum. The presence or absence of an optical counterpart was not part of the flagging process as we did not want to introduce bias into the \hi catalogue stemming from the limitations of the available optical catalogues. 
The flags are as follows:
\begin{itemize}
\item 0 - certain detection without any additional flags;
\item 1 - uncertain detections. Those are usually low-S/N sources or sources detected close to a region with known Radio-Frequency Interference (RFI). All other flags indicate certain detections;
\item 2 - indicates \hi detections with more than one optical galaxy likely contributing to the measured signal;
\item 3 - the source is at the edge of the cube and part of the flux is likely missed;
\item 4 - indicates sources that are likely resolved by the $3.4\arcmin$ beam of Arecibo, and therefore the tabulated total flux and M$_{\text{\sc H\,i}}$ should be treated as lower limits.
\end{itemize}

The flags are listed in the online tables. A breakdown of the catalogue by method of detection (visual/automatic) and flag is given in Table \ref{table_sources}. The visual inspection was based on an atlas presenting the sources and their parameters. Example of this atlas, showing sources with a range of S/N and flag, is shown in Appendix \ref{atlas}. The full atlas is available for download \footnote{\url{https://www.dropbox.com/s/xffshfom5h0m3f0/AGES_A1367_atlas.tar.gz?dl=0}} (0.5 GB).

%\href{https://www.dropbox.com/s/xffshfom5h0m3f0/AGES_A1367_atlas.tar.gz?dl=0}{here} (0.5 GB).

Following the above search for detections, an \hi spectrum was extracted at the position of every SDSS spectroscopic target within the AGES volume. Those were visually inspected but did not yield any additional detections.

In previous works, such as \citet{cortese2008}, we used Arecibo to perform follow-up observations to verify uncertain detections. During the course of this analysis, this became impossible; however, we obtained 5 hours of observing time on the Five-hundred-metre Aperture Spherical Telescope (FAST) to re-observe uncertain detections. As we are concerned with the large-scale distribution of sources rather than individual detections, we leave the analysis of these observations to a subsequent paper and omit uncertain detections from further analysis.
\\

% Table with the sources by detection method and flag
\begin{table}
\caption{Breakdown of the \hi catalogue by method of detection and flag}
\label{table_sources}
\begin{tabular}{c|ccccc|c}
Extraction & \multicolumn{5}{|c|}{Flag} & \multirow{2}{*}{Total}\\
 \cline{2-6}
Method & 0 & 1 & 2 & 3 & 4 &  \\
 \hline
 \hline
FRELLED & 358 & 9 & 19 & 5 & 21\tablefootmark{a} & 412 \\
GLADoS & 9 & 35 & 0 & 1 & 0 & 45 \\
 \hline
Total & 367 & 44 & 19 & 6 & 21 & 457 \\
\end{tabular}
\tablefoot{
\tablefoottext{a}{9 objects have flag 4 in addition to their primary flag. Only primary flags are in the table}}
\end{table}

\subsection{Optical counterparts}
We searched for optical counterparts of the \hi detections among the galaxies with SDSS spectroscopic redshifts. The search region around the centre of each \hi detection was 1.7\arcmin (FWHM/2) projected on the sky, and in velocity range $\left|V_{\text{\sc H\,i}}-V_{opt}\right| \leq$ W$_{20}$/2 + $\Delta$W$_{20}$, where W$_{20}$ is the velocity width of the line profile at 20\% of the peak intensity and $\Delta$ W$_{20}$ its uncertainty. Similar criteria were used in all previous AGES papers. This method confirmed the optical counterparts of 322 \hi detections, with a median distance between the centroid of the \hi emission and the centre of the optical counterpart of 12.5 \arcsec and velocity difference of 7 km s$^{-1}$. When more than one galaxy satisfies those criteria, the closest one on the sky is listed as the primary optical counterpart. Those galaxies have flag = 2. Many of the resolved galaxies with flag = 4 also have more than one optical counterpart, and therefore the total number of \hi-detected galaxies in this field is likely significantly higher than 457. Additional searches for optical counterparts were carried out in the NASA/IPAC Extragalactic Database (NED), yielding a further 13 objects with spectroscopic redshift, bringing the total to 335. In addition, the online table lists the most likely optical counterpart for 106 of the 122 remaining \hi sources. Those galaxies lack optical redshifts but lie within the \hi beam. We could not find optical counterparts for 16 of our \hi detections. Those are almost all low S/N detections, 13 of which are flagged as uncertain (flag = 1).

Some additional properties of the optical counterparts of the \hi detections needed for the analysis presented in the following sections were taken from publicly available data sets. The stellar mass is from MPA-JHU value-added catalogue based on SDSS DR8 and assume a \citet{Chabrier2003} initial mass function. The details of the methods used to derive the stellar mass are given in \citet{Kauffmann2003}, \citet{Tremonti2004}, and \citet{Brinchmann2004}. The catalogue provides stellar mass estimates for 296 of the 335 optical galaxies which we detect in \hi. Where relevant, this is the sample used in the following analysis. The r-band size of the stellar discs is from SDSS \citep{Ahumada2020}.

\subsection{Optical sample and upper limits on \Mhi for non-detections}\label{section:up_lims}
SDSS provides, through the MPA-JHU catalogue, the precise positions, redshifts, and stellar masses for 830 galaxies within the AGES survey volume. We use this as our optical sample in the following analysis. Of those, 296 objects are directly detected in \hi and catalogued. For the remaining 534, we need to estimate the maximum amount of \hi that could be hosted by them without being detected. We base our estimate on the rms noise measured per pixel in the data cube. Assuming that those galaxies are not spatially resolved, we need to estimate the velocity width of their signal. We do that using the i-band Tully-Fisher relation \citep[TF, ][]{Tully_Fisher1977} published by \citet{Masters2006} and SDSS i-band Petrosian magnitudes. A two sample Kolmogorov-Smirnov test between the profile widths of non-detections, derived as described above, and the measured W$_{50}$ of \hi-detected galaxies returned a p-value of 4 $\times$ 10$^{-6}$, indicating that the two samples can be drawn from the same parent distribution.

\section{Results} \label{section:Results}
%----------------------------------------------------------------- 
   \begin{figure}
   \centering
   \includegraphics[width=\hsize]{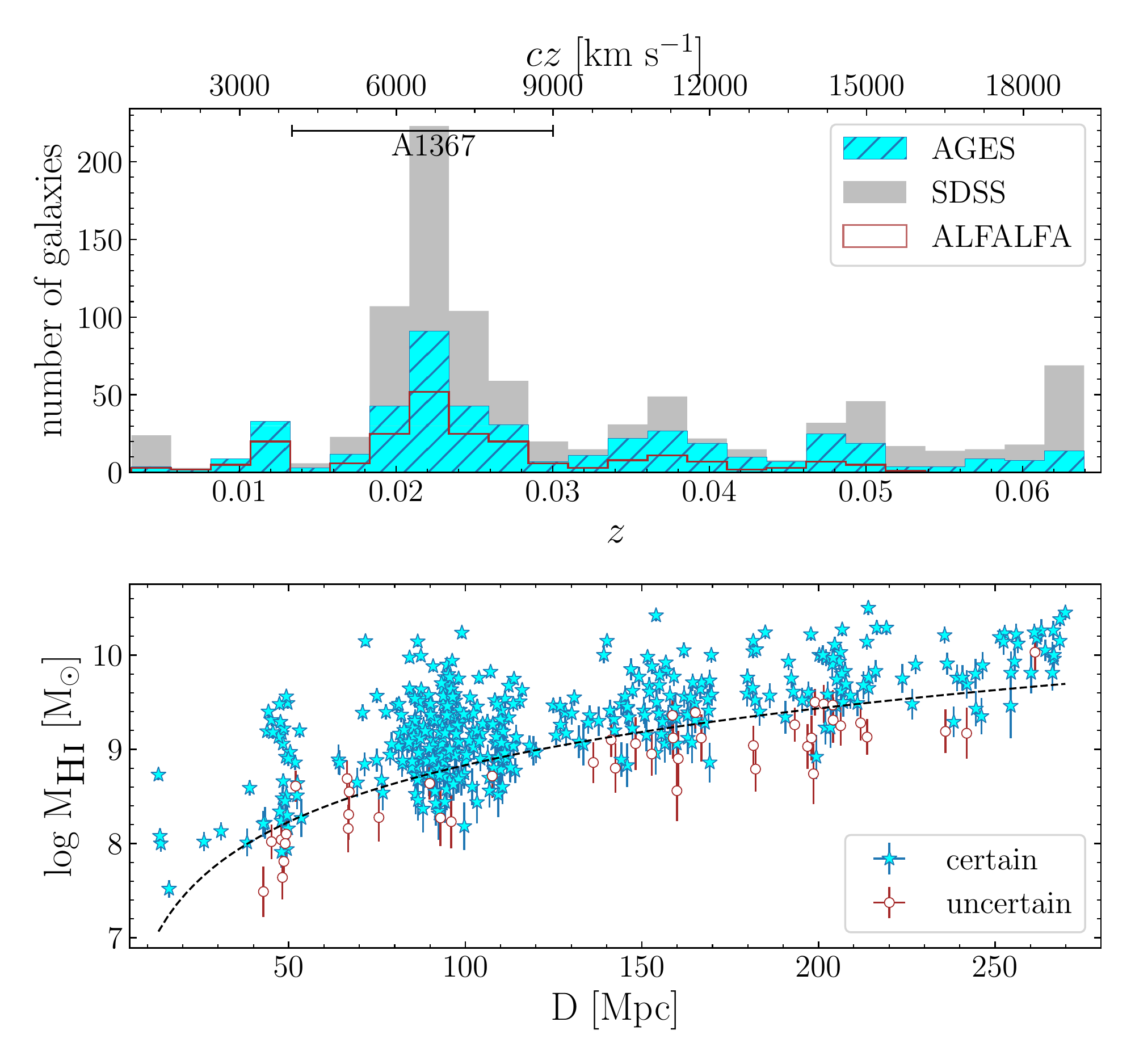}
      \caption{\textit{Top:} Redshift distribution of the AGES \hi detections (cyan hatched histogram) compared to SDSS spectroscopic galaxies (solid grey histogram), and ALFALFA detections (empty brown histogram) within the AGES volume. The velocity range of A1367 is indicated with the horizontal bar. \textit{Bottom:} Distribution of M$_{\text{\sc H\,i}}$ as a function of distance to the source, assuming Hubble flow distances. The black, dashed line shows the mass of a galaxy with velocity width of 200~km~s$^{-1}$ detected at 6.5~$\sigma$. Certain and uncertain detections are shown with blue stars and brown circles, respectively. \href{https://github.com/deshev/AGES-XI/blob/main/plot_z_distr.py}{\includegraphics[height=3mm]{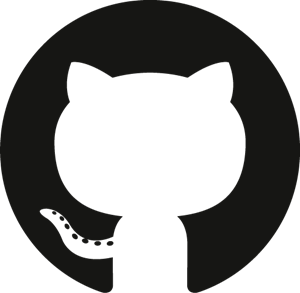}}}
         \label{z_distr}
   \end{figure}
%-----------------------------------------------------------------
%\footnotetext{Throughout this article, clicking on the GitHub logo will open the Python script used for executing the calculation or producing the figure being discussed.}
%----------------------------------------------------------------- 
   \begin{figure}
   \centering
   \includegraphics[width=\hsize]{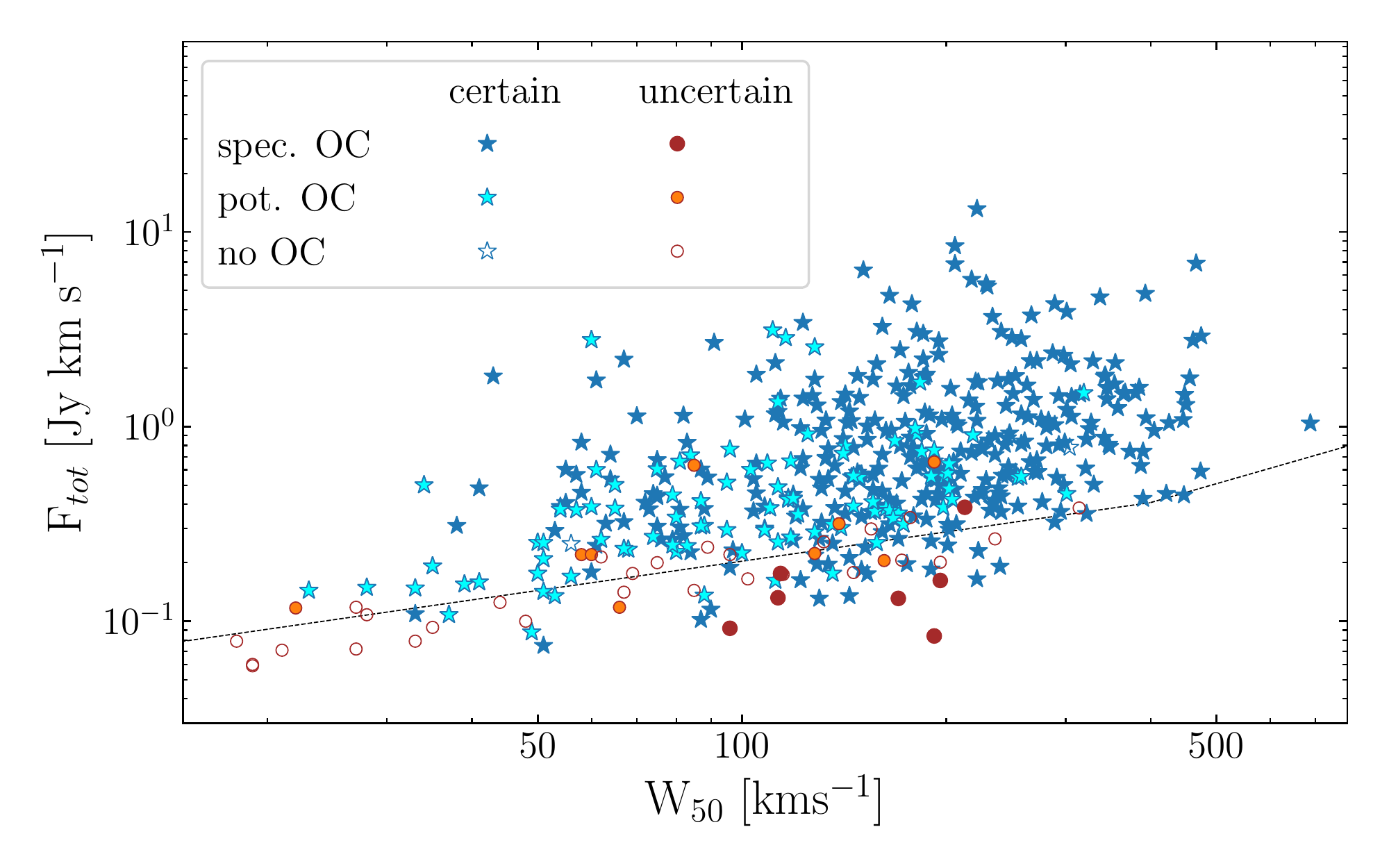}
      \caption{Distribution of total flux in the 21 cm line as a function of line width for the certain detections (stars) and uncertain ones (circles). Filled and empty symbols show detections with and without an optical counterpart (OC), respectively, with the different shade fill segregating the spectroscopically confirmed and the potential optical counterparts. The dashed black line shows the ALFALFA reliability limit \citep{saintonge2007} at 6.5$\sigma$. \href{https://github.com/deshev/AGES-XI/blob/main/plot_flux_distr.py}{\includegraphics[height=3mm]{github-logo.png}}}
         \label{fl_distr}
   \end{figure}
%-----------------------------------------------------------------
%----------------------------------------------------------------- 
   \begin{figure}
   \centering
   \includegraphics[width=\hsize]{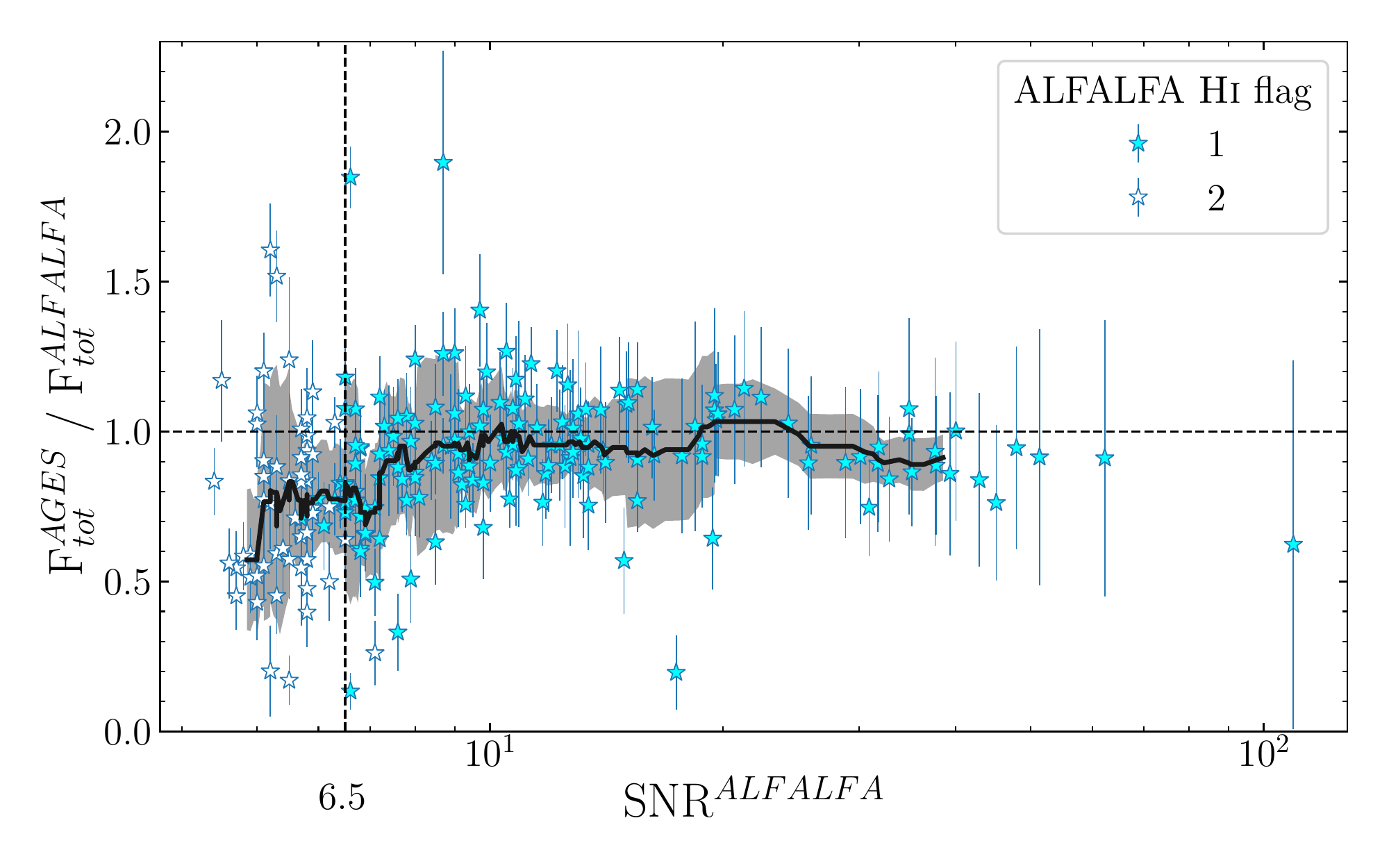}
      \caption{Comparison between the measured total line flux of \hi sources common to both ALFALFA and AGES surveys. The flux ratio is plotted as a function of the ALFALFA S/N. The black curve and shaded region represent a running median and $\pm$1$\sigma$. The detection reliability limit is indicated with the vertical dashed line.  \href{https://github.com/deshev/AGES-XI/blob/main/plot_comparison.py}{\includegraphics[height=3mm]{github-logo.png}}}
         \label{flux_comparison}
   \end{figure}
%-----------------------------------------------------------------

The complete list of all \hi detections and their measured properties- which are used in the following analysis- is given in the online tables. The list with the optical counterparts of the \hi detections and their properties is also provided.

The distribution of measured total \hi mass (M$_{\text{\sc H\,i}}$) and overall distribution of all AGES detections as a function of redshift $z$ are shown in Fig.~\ref{z_distr} in comparison to the distribution of SDSS spectroscopic galaxies and ALFALFA detections \citep{Haynes2018} within the AGES survey volume. The AGES catalogue contains 457 \hi detections, 413 of which are considered certain, almost doubling the source count compared to ALFALFA (212). The density of \hi detections follows the distribution of galaxy concentrations in velocity as traced by SDSS from optical spectroscopy. Also visible is the relative absence of \hi detections in ALFALFA at high-$z$. 

We have a number of \hi detections for which there is no spectroscopically confirmed optical counterpart (even if most have an obvious candidate, which are also listed in the online tables),  and there are many optical galaxies without a corresponding \hi detection. At the high-$z$ end, the AGES survey contains $\sim$~50\% fewer galaxies than SDSS. This fraction rises steadily towards the low-$z$ end of the survey, with a noticeable decrease in \hi detections at the velocity range of A1367. The growth of the fraction of \hi-detected galaxies towards lower redshift is explained by the higher \hi mass sensitivity shown in the bottom panel of Fig.~\ref{z_distr}. The black dashed curve shows the expected M$_{\text{\sc H\,i}}$ of a galaxy with W$_{20}$ = 200 km s$^{-1}$ detected at 6.5 $\sigma$. The fraction of optical galaxies detected in \hi is analysed in detail in Section \ref{section:Discussion}. The percentage of sources below the 6.5$\sigma$ limit from the five separate flags listed in Table \ref{table_sources} is 10, 77, 5, 0, 0, respectively (qualitatively this information is also conveyed in Fig.\ref{fl_distr}).\\

Figure \ref{fl_distr} shows the distribution of the 457 \hi detections in the plane defined by their total flux in the line and the width of the line measured at 50 \% of the maximum of the flux (W$_{50}$). The sources are split into certain and uncertain ones, according to their detection flag. Also indicated in the table is the presence or absence of a spectroscopically confirmed or photometric optical counterpart. The dashed black line shows the distribution of galaxies at constant S/N = 6.5 following \cite{saintonge2007} and \cite{cortese2008}. As concluded by the latter, this is a reliable limit for separating the certain and uncertain detections. Of all AGES detections, 13\% (72) are below this line, 47\% (34) of which are flagged as uncertain. The median uncertainty in the total flux in the line for all AGES sources is $\Delta$F $\sim$ 0.1 Jy km s$^{-1}$. Within one $\Delta$F above and below the 6.5 $\sigma$ line, the fraction of uncertain detections changes from 0 to $>$ 50\%. As pointed out by \cite{cortese2008}, the lack of sources at the high-velocity-width, low-flux regime is a result of the fact that weak sources have a better chance of being detected if their flux is distributed over fewer resolution elements, which gives preference to face-on oriented galaxies. This is reflected in the slope of the dashed black line.\\

Figure \ref{flux_comparison} shows the ratio of the total flux in the line measured from the AGES and ALFALFA surveys for the 212 \hi sources common to both surveys. This is presented as a function of the S/N as measured by the ALFALFA survey and published by \citet{Haynes2018}. The error bars are a quadrature sum between the tabulated uncertainties and those published by \citet{Haynes2018}. The thick black curve and shaded area represent a running median over a window of 15 sources ($\sqrt{\text{N}}$ = 15, where N = 212 is the total number of sources common to the two catalogues). The vertical dashed line shows the reliability limit for detecting an \hi source in the ALFALFA survey. AFLALFA offers two quality flags on the published \hi flux, based on S/N and the availability of optical counterpart, which are indicated with different symbols. As a result of the greater integration time employed by AGES, none of these sources are flagged as uncertain. The increased scatter at low S/N is expected \citep{Haynes2011}. The systematic drop in the flux ratio at low S/N can be explained as a manifestation of Eddington bias \citep{Eddington1913} where faint sources close to and below the detection limit are only detected if their flux is artificially boosted by fortuitous noise distribution. AGES detects those same sources at higher S/N and shows lower, more accurate flux.

\subsection{Local galaxy density \href{https://github.com/deshev/AGES-XI/blob/main/density_estimator.py}{\includegraphics[height=3mm]{github-logo.png}}}\label{section:gal_density}
%----------------------------------------------------------------- 
   \begin{figure}
   \centering
   \includegraphics[width=\hsize]{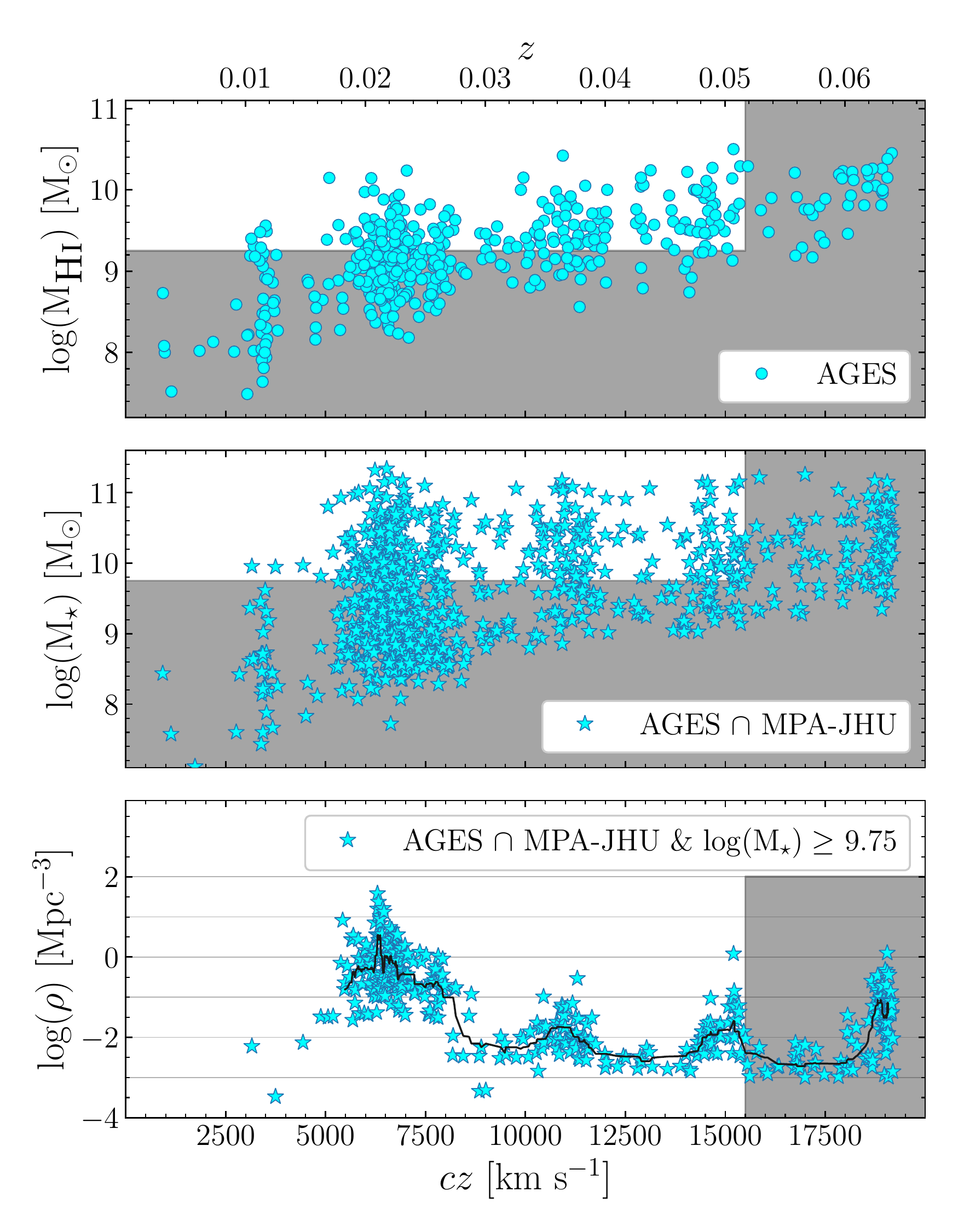}
      \caption{Distribution of \hi mass for all AGES detections(top) and stellar mass for all optical galaxies in the intersection between AGES and MPA-JHU (middle) as a function of recession velocity. The shaded regions indicate the areas not complete in \Mhi and M$_{\star}$, respectively. The bottom panel shows the local galaxy density for the mass-complete part of the middle sample. The black line shows the running median. \href{https://github.com/deshev/AGES-XI/blob/main/plot_mass_distributions.py}{\includegraphics[height=3mm]{github-logo.png}}}
         \label{mass_distributions}
   \end{figure}
%----------------------------------------------------------------- 
The top two panels of Fig. \ref{mass_distributions} show the distribution of \Mhi for all AGES detections and the stellar mass for all galaxies in the MPA-JHU catalogue within the AGES volume. The white areas on those two panels define the \hi mass and stellar mass complete samples, used below to calculate the local density of galaxies. The limits of M$_{\star}$ $\leq$ 9.75 M$_{\odot}$ and \Mhi $\leq$ 9.25 M$_{\odot}$ are defined as the peaks of the corresponding mass distributions in a narrow redshift bin centred at 15500 km s$^{-1}$. In the following analysis, we did not use the higher redshift part of the survey due to the presence of RFI.

%----------------------------------------------------------------- 
\begin{landscape}
\begin{figure}
\resizebox{24.5cm}{!}
{\centering\includegraphics{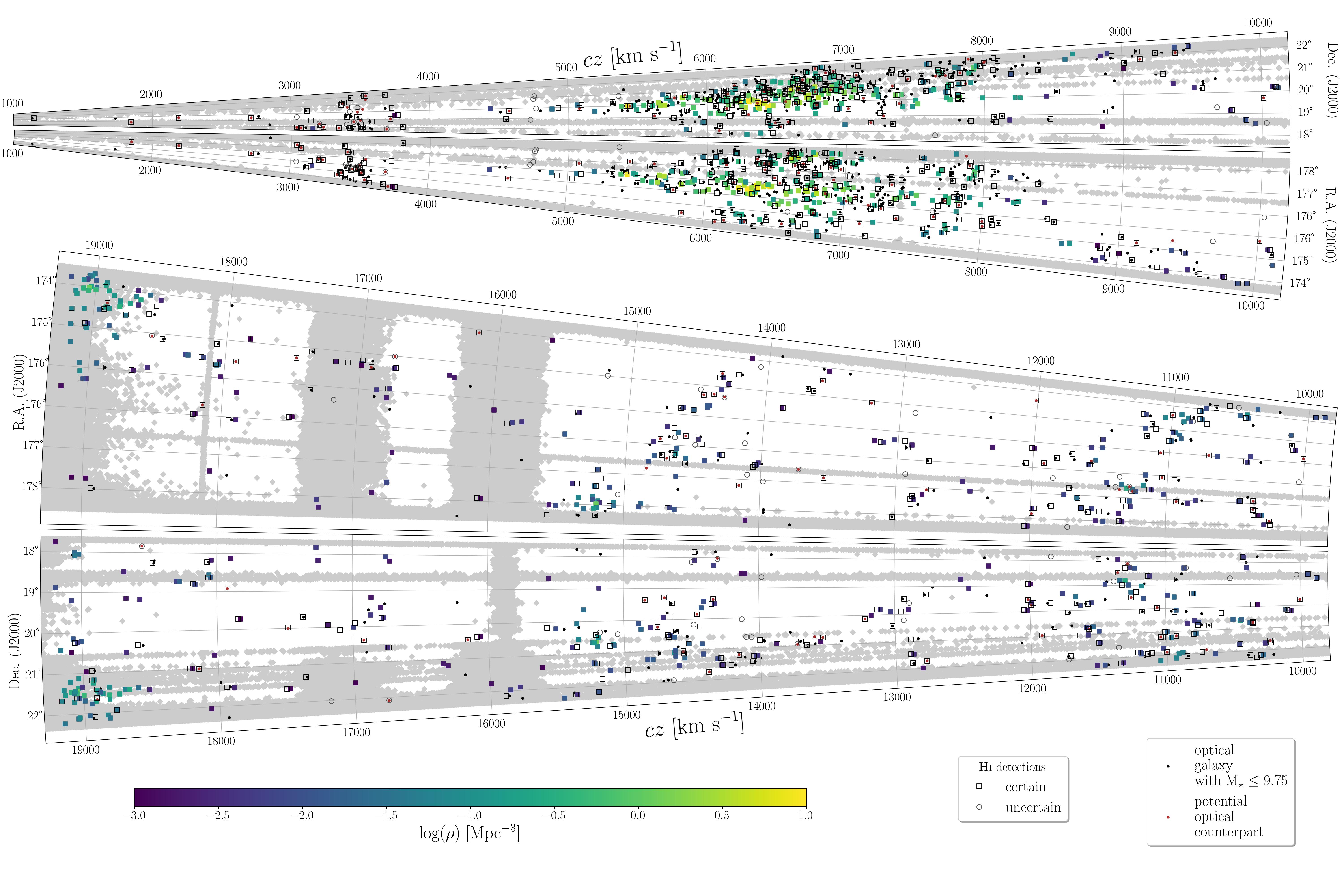}}
\caption{ Polar plot showing the distribution of the \hi-detected and optical galaxies in the R.A. - $cz$ and Dec. - $cz$ planes. The $z$ range of the survey is split in two for clarity. Large squares of different colour show galaxies with spectroscopic redshift from SDSS and with stellar mass M$_{\star}\geq$ 9.75 M$_{\odot}$ with the colour indicating their local density according to the colour bar. Small black points show optical galaxies with stellar mass below this limit. Small brown points mark potential optical counterparts without optical redshift. Certain \hi detections are shown with empty, black squares, uncertain ones with empty, dark-grey circles. The light-grey shaded areas of the map show areas where the ability to detect \hi signal is reduced (see text for details). The figure is meant to be viewed on a (large) screen. \href{https://github.com/deshev/AGES-XI/blob/main/make_polar_plot.py}{\includegraphics[height=3mm]{github-logo.png}}}
\label{polar_plot}
\end{figure}
\end{landscape}
%----------------------------------------------------------------- 

The local galaxy density is calculated from the volume of a Voronoi cell centred on every galaxy and with a size equal to half the distance to the nearest galaxy in all directions, similar to the Voronoi-Delaunay method \citep{Voronoi1908, Delaunay1934, Weygaert1994, Marinoni2002, Pereira2017}. This is calculated in a 3D space of comoving physical distances for the mass-complete sample, and is therefore only available for galaxies with log M$_{\star} \geq$ 9.75 M$_{\odot}$. In this case, we used the redshift to calculate the distance to the objects, including cluster galaxies. This allows us to calculate and compare the local galaxy density in the same way throughout the surveyed volume. Estimating distance from redshift for virialised structures is affected by the Fingers-of-God \citep[FoG effect, ][]{Jackson1972, Tully1978} and Kaiser \citep{Kaiser1987} effects. The two work in opposite directions, and in general the Kaiser effect is expected to be weaker than FoG \citep{Kuutma2017}. By comparing the distribution on the sky of SDSS members of A1367 with their distribution along the line of sight, both approximated with a Gaussian, we determined that the cluster is stretched along the latter by a factor of 3.7. Therefore, we reduce all cluster-centric distances for A1367 members along the line of sight by this amount. We note that this does not completely remove the FoG effect, which is impossible without precise redshift independent distances; instead it minimises the artificial lowering of 3D densities in the cluster.

The calculation of the Voronoi tesselations and their volumes was done in Python using \textsc{SciPy} packages \textsc{Voronoi} and \textsc{Delaunay}. The distribution of local density is shown in the bottom panel of Fig. \ref{mass_distributions}, and further analysed in Section \ref{section:Discussion}.

\subsection{Observed phase space view of the surveyed volume}

Figure \ref{polar_plot} presents the \hi detections in the context of the large-scale structure in which they are embedded, where the surveyed volume is shown with its true axis ratio. The recession velocity range has been split in two for clarity. The volume is represented in two planes, collapsing the Dec or R.A. ranges, respectively. The figure shows the local density of optical galaxies from the stellar-mass-complete sample, colour coded according to the colour bar. The density clearly shows the position of the cluster A1367, centred at around 6500 km s$^{-1}$, but also a number of other structures with a range of densities. The potential optical counterparts are shown in Fig.\ref{polar_plot} as small brown points plotted at their measured R.A. or Dec, and the recession velocity of the \hi signal. The optical galaxies with stellar mass below the completeness limit, for which local density is not available, are shown as small black points. The figure also shows the areas of the survey affected by RFI or with increased noise levels at the edges of the data cube. Those areas are marked with light-grey diamonds and show R.A. or Dec columns with rms noise deviating from the mean rms of the cube by more than 4.5 $\sigma$. We note that in some cases (particularly with the presence of RFI), this does not need to affect the entire column, hence the presence of some \hi detections in an area marked in grey. However, the plot clearly shows entire redshift bands where \hi detections are almost completely absent due to RFI, namely around 16000 and 17000 km s$^{-1}$, but also a narrow stripe at around 7300 km s$^{-1}$ running through the cluster.

\subsection{\hi deficiency \href{https://github.com/deshev/AGES-XI/blob/main/calc_deficiency.py}{\includegraphics[height=3mm]{github-logo.png}}}\label{section:deficiency_results}
%----------------------------------------------------------------- 
   \begin{figure}
   \centering
   \includegraphics[width=\hsize]{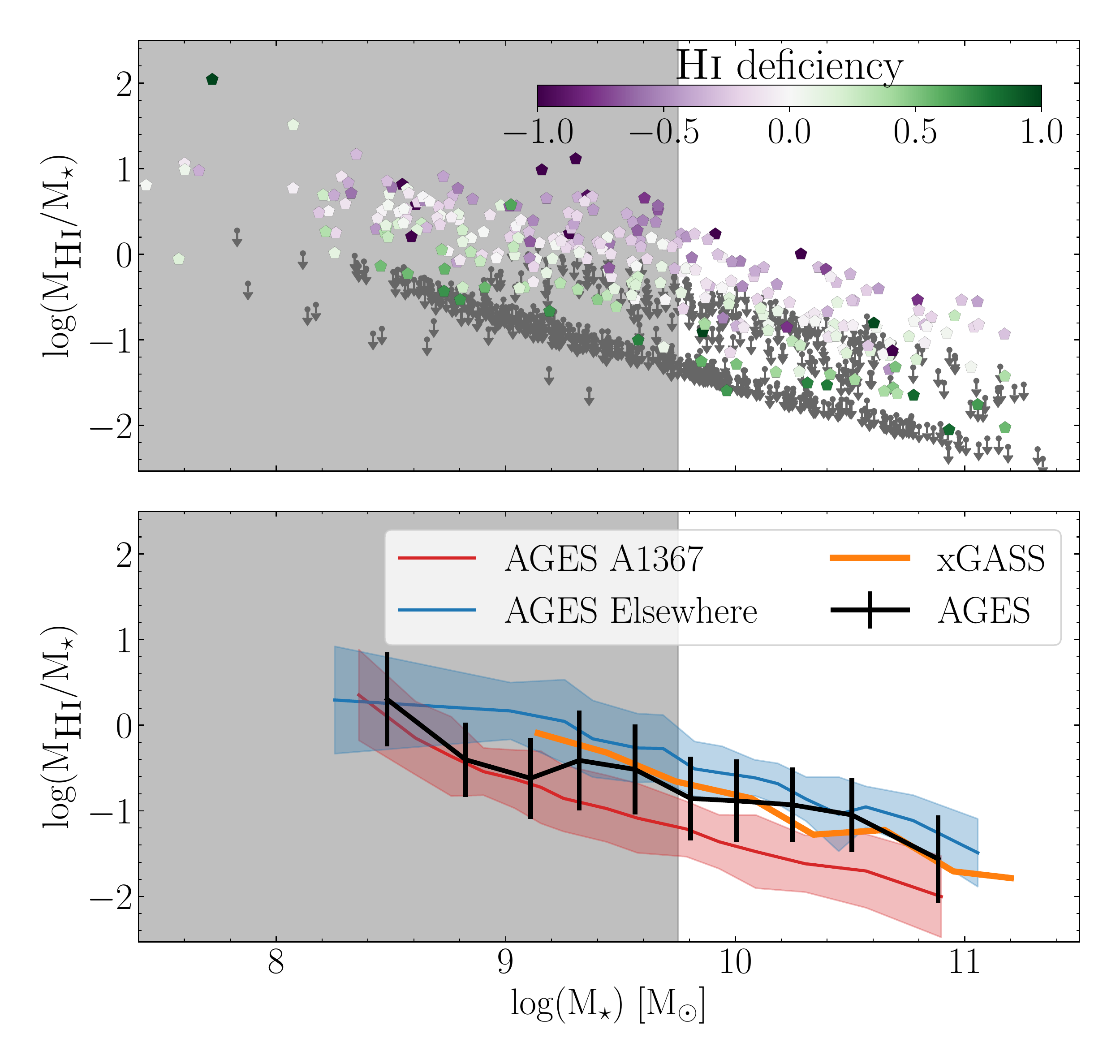}
      \caption{Ratio of total mass in atomic hydrogen to total stellar mass for all AGES detections with known optical counterparts (pentagons) and upper limits for non-detections (grey, down-pointing arrows). The pentagons are colour coded according to their \hi deficiency. The thick black line shows the median of all AGES galaxies. The orange solid line shows the median values for xGASS galaxies \citep{Catinella2018}. The grey shaded region encompasses the data excluded from the stellar-mass-limited sample. The red and blue curves and shaded regions show the median relations for AGES galaxies that are members of the A1367 cluster and non-members, respectively. \href{https://github.com/deshev/AGES-XI/blob/main/plot_mhimstar.py}{\includegraphics[height=3mm]{github-logo.png}}}
         \label{mhimstar}
   \end{figure}
%-----------------------------------------------------------------

It is instructive to compare the amount of \hi-detected in individual galaxies with a baseline average of the entire population in order to assess the presence of any physical mechanisms that can alter the gas content of galaxies, be it through gas removal or through prevention of accretion. We employed a formula published by \citet{Haynes&Giovanelli1984} which links the total gas mass to the size of the optical disc of the host galaxy. 

\begin{equation}
DEF_{\text{\sc H\,i}} = \text{log}_{10}\left(\frac{\text{\Mhi}^{expected}}{M_{\odot}}\right) - \text{log}_{10}\left(\frac{\text{\Mhi}^{observed}}{M_{\odot}}\right)
\end{equation}

This relies on the relatively tight correlation between the size of the \hi disc and its total mass on one hand and the size of the optical disc on the other. \citet{Haynes&Giovanelli1984} provided two separate formulas to be used with galaxies with early- and late-type morphologies, respectively. We found 149 galaxies among the \hi-detected sample with optical counterparts for which the Galaxy Zoo project contains morphologies determined with certainty \citep{Lintott2011}. Of those, only 7 were classified as elliptical. We therefore decided to only use the formula for spiral galaxies when calculating \hi deficiencies. In particular, we took the Petrosian radius, which contains 90\% of their r-band luminosity. In order to bring the Petrosian radius offered by the SDSS pipeline into agreement with the size estimates used by \citet{Haynes&Giovanelli1984}, we applied a correction calculated by comparing petroR90 with the radii of galaxies measured at surface brightness of 25 magnitudes arcsec$^{-1}$ (D$_{25}$) by the GOLDmine project \citep{Gavazzi2003a}. We found 47 galaxies in common between the Goldmine data base and the catalogue of optical counterparts of the \hi detections presented here. The ratios of the two size estimates were normally distributed with a mean of 1.4 and standard deviation of 0.3. Thus, the expected \hi mass of all optical galaxies is: M$_{\text{\sc H\,i}}^{expected}$ = $7.195 + 0.851 \times log{(2 \times PR90 \times 1.4)^{2}}$, where PR90 is the Petrosian radius in kiloparsecs. The deficiency is defined as the difference in the logarithms of the expected and observed M$_{\text{\sc H\,i}}$. 

The results are presented in Fig. \ref{mhimstar} and analysed further in Section \ref{section:defficiency_discussion}. Figure \ref{mhimstar} shows all \hi detections with optical counterparts demonstrating the known, relatively tight relation between the \hi mass fraction, M$_{\text{\sc H\,i}}$/M$_{\star}$, and M$_{\star}$. All points are colour coded according to their \hi deficiency calculated as described above. The small grey points and down-pointing arrows show the upper limit on gas fraction for galaxies not directly detected  in \hi, calculated as explained in Section \ref{section:up_lims}. Also shown is the relation from the xGASS survey, published by \citet{Catinella2018}, which is to be compared with the median of the AGES data shown with the black line and error bars.

\section{Discussion} \label{section:Discussion}
The part of the AGES survey presented here has an on-the-sky extent of 5 $\times$ 4 degrees, which corresponds to 18.6 $\times$ 23.2 comoving megaparsecs, at the distant end of the survey. In contrast, the distance between the nearest and most distant \hi detections presented in this work is 257 Mpc. This large distance defines the significant variation in sensitivity over the surveyed volume. For example, with the very stable noise level across the volume, a single resolution element with 3 $\sigma$ signal would be equivalent to log M$_{\text{\sc H\,i}}$ = 6.0 M$_{\odot}$ at the near end of the survey and 8.6 at the far end. This large difference in sensitivity is expressed with the black dashed curve on Fig. \ref{z_distr}. For this reason, we analysed the surveyed volume in three bins which also represent different environments. The low-$z$ foreground part of the cluster A1367 spans up to $cz$ = 4000 km s$^{-1}$ and has a total volume of 361 Mpc$^{3}$ and average 6.5 $\sigma$ sensitivity for a galaxy with 200 km s$^{-1}$ velocity profile of log M$_{\text{\sc H\,i}}$ = 7.8 M$_{\odot}$. The second part contains the cluster A1367 and is defined as the volume between 4000 $\leq cz \leq$ 9000 km s$^{-1}$, with a total volume of 3719 Mpc$^{3}$ and an average sensitivity of log M$_{\text{\sc H\,i}}$ = 8.7 M$_{\odot}$. The last part lies at $cz$ > 9000 km s$^{-1}$ and has a total volume of 39940 Mpc$^{3}$ and an average sensitivity of log M$_{\text{\sc H\,i}}$ = 9.4 M$_{\odot}$.

\subsection{\hi below the detection threshold}\label{section:stacking}
%----------------------------------------------------------------- 
   \begin{figure*}
   \centering
   \includegraphics[width=\hsize]{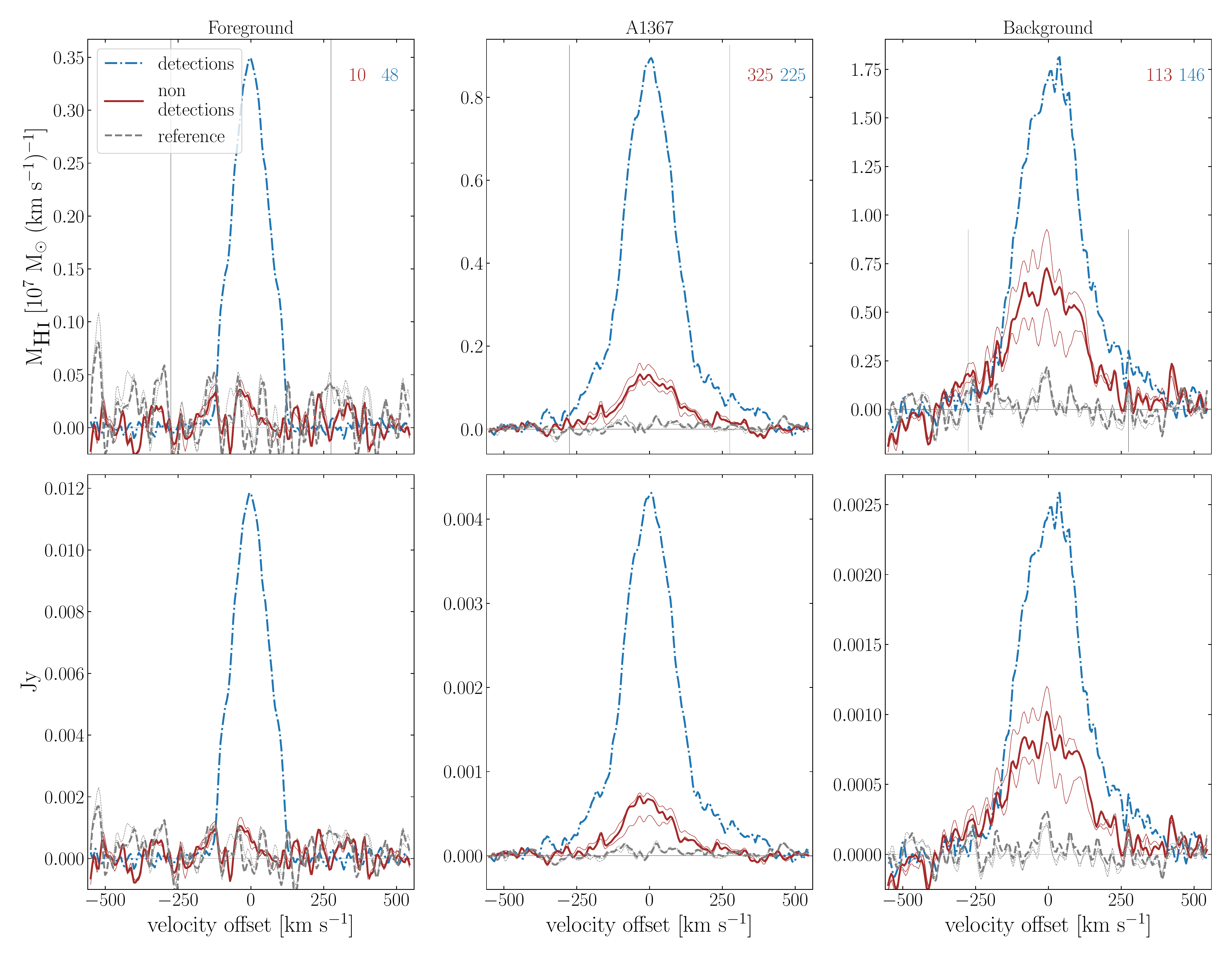}
      \caption{\hi signal present in the data below the detection threshold (brown curves) recovered by stacking spectra from galaxies in A1367, those infront of and behind it. The thick lines show the median and the thin ones the 95\% confidence interval of the distribution of total M$_{\text{\sc H\,i}}$ in the stacked spectra, from jackknife resampling. For comparison, the dash-dotted blue lines shows the average spectrum of all individually detected galaxies. The dashed grey curves show the signal distribution in a reference stack (see text for details). The vertical grey lines show the velocity range over which the curves are integrated. The total number of galaxies in each stack is indicated at the top, coloured in blue for detections and red for non-detections. The bottom row shows the same spectra but in flux units.\href{https://github.com/deshev/AGES-XI/blob/main/undetected_stack.py}{\includegraphics[height=3mm]{github-logo.png}}}
         \label{stack}
   \end{figure*}
%----------------------------------------------------------------- 
%-------------------------------------------------------------
%
\begin{table}
\caption{Average \hi mass in Fig. \ref{stack} in units of 10$^{8}$ M$_{\odot}$ }             % title of Table
\label{table_stack}      % is used to refer this table in the text
\centering                          % used for centering table
\begin{tabular}{c c c c}        % centered columns (4 columns)
\hline\hline                 % inserts double horizontal lines
 & Foreground & A1367 & Background \\    % table heading 
\hline                        % inserts single horizontal line
Detected & 7.5 &  19.8 &  47.1 \\      % inserting body of the table
Non-detected & 0.41$^{+0.16}_{-0.15}$ & 3.19$^{+0.64}_{-0.63}$ & 19.9$^{+4.89}_{-5.15}$ \\
\hline                                   %inserts single line
\end{tabular}
\end{table}
%
%-------------------------------------------------------------

Taking the reliability threshold shown in the bottom panel of Fig. \ref{z_distr}, namely 6.5 $\sigma$ for an \hi profile over 200 km s$^{-1}$, means that we are sensitive to M$_{\text{\sc H\,i}}$ $\leq$ 9 and $\leq$ 8 over 10\% and 4\% of the total surveyed volume of 44021 Mpc$^{3}$, respectively. The 'by eye' strategy for source finding that we adopted here, plus the occasionally more favourable velocity distribution of the \hi signal, allows us to find a significant number of detections lying below this threshold - $\sim$ 16\% of all detections, with approximately half of those considered certain. Despite this, we expect that some amount of \hi emission is present in the data cube at very low S/N and is therefore not catalogued. Traditionally, the \hi community uses spectral stacking to recover such emission lying at low S/N \citep{Fabello2012, Gereb2015, Hu2021}. This method uses the availability of known on-the-sky positions and the redshift of galaxies (usually from optical spectroscopy) to extract their \hi spectra. Averaging these spectra together will average out the Gaussian noise while adding the signal to produce one \hi detection representing all the otherwise undetected galaxies. 

Of the 830 galaxies in the optical sample, 296 are individually detected and catalogued, leaving 534 available for stacking. Figure \ref{stack} shows the results from such a stacking experiment. For every such galaxy, we extracted a spectrum from the \hi data cube in the same way as the spectra of individual detections were extracted. Those spectra were then converted to units of M$_{\text{\sc H\,i}}$ / km s$^{-1}$ using the formula listed in Section \ref{section:Results}. Following that, weights were calculated based on the rms noise in the outer parts of the spectra, which are presumed not to contain any signal.

The galaxies were split into three bins of recession velocity: A1367 members, foreground, and background. Additionally, all galaxies lying in the RFI affected regions at $cz \geq$ 15500 km s$^{-1}$ were removed from the background bin. This left a total of 10, 325, and 113 galaxies in the three bins, respectively. Subsequently, 10000 iterations were done with jackknife resampling where 90\% of the available spectra were randomly selected, without replacement, and stacked. This was done to detect the presence of highly deviant spectra. The total M$_{\text{\sc H\,i}}$ in each stacked spectrum was integrated in the velocity range between the two vertical grey lines in Fig. \ref{stack}. The thick lines show the spectra that gave the median of the distribution of M$_{\text{\sc H\,i}}$ and the thin ones show the 95\% confidence interval. 

The average \hi content of the galaxies in each stack as measured from these spectra is given in Table \ref{table_stack}. The observed increase in the average \hi mass in the directly detected galaxies and in the stack of non-detections is likely driven by the sensitivity curve shown in Fig. \ref{z_distr} which, as the redshift increases, removes low-mass galaxies from the stack with detections and moves them into the stack with non-detections. To confirm the presence of the signal in the stacked spectra, we also show a reference stack in grey in each panel; these are extracted at the positions of the optical galaxies but with randomised redshifts and stacked in the same way.
The bottom row of panels on Fig. \ref{stack} shows the same stacked spectra but in flux units, thus avoiding the distance$^{2}$ factor used in the conversion of flux to \hi mass.

A comparison between the stacked and reference spectra in the foreground bin shows that the measured \hi mass from integrating the spectra is consistent with a non-detection. The small number of galaxies in this stack prevents us from reaching sufficiently low noise levels to draw any conclusions on the \hi content of galaxies with \Mhi below 10$^{7}$ M$_{\odot}$. 

Previous studies aimed at detecting galaxies with very low gas content \citep[e.g.][]{Popping_PhD} show that objects with \hi masses below 10$^{7}$ M$_{\odot}$ are very rare. On the contrary, \citet{Jones2018} find that the \hi mass function does not show a decline until its measured limit of \Mhi = 10$^{6.5}$ M$_{\odot}$. The reason for this may lie in a combination of the semi-constant column density distribution of \hi within the galaxies, expressed in the mass -- size relation for \hi discs \citep{Prochaska2009, Wang2016MSrelation}, and the higher susceptibility of \hi to photoionisation at low column densities \citep{Schaye2001}. Interferometric observations confirm the constancy of the relation down to log M$_{\text{\sc H\,i}}$ $\sim$ 7 \citep{Wang2016MSrelation}. Beam dilution, as a consequence, also plays a role, especially for single-dish surveys, as this relation drives the sizes of the \hi discs well below the size of the beam.

\subsection{What goes into \hi deficiency}\label{section:defficiency_discussion}
%----------------------------------------------------------------- 
   \begin{figure}
   \centering
   \includegraphics[width=\hsize]{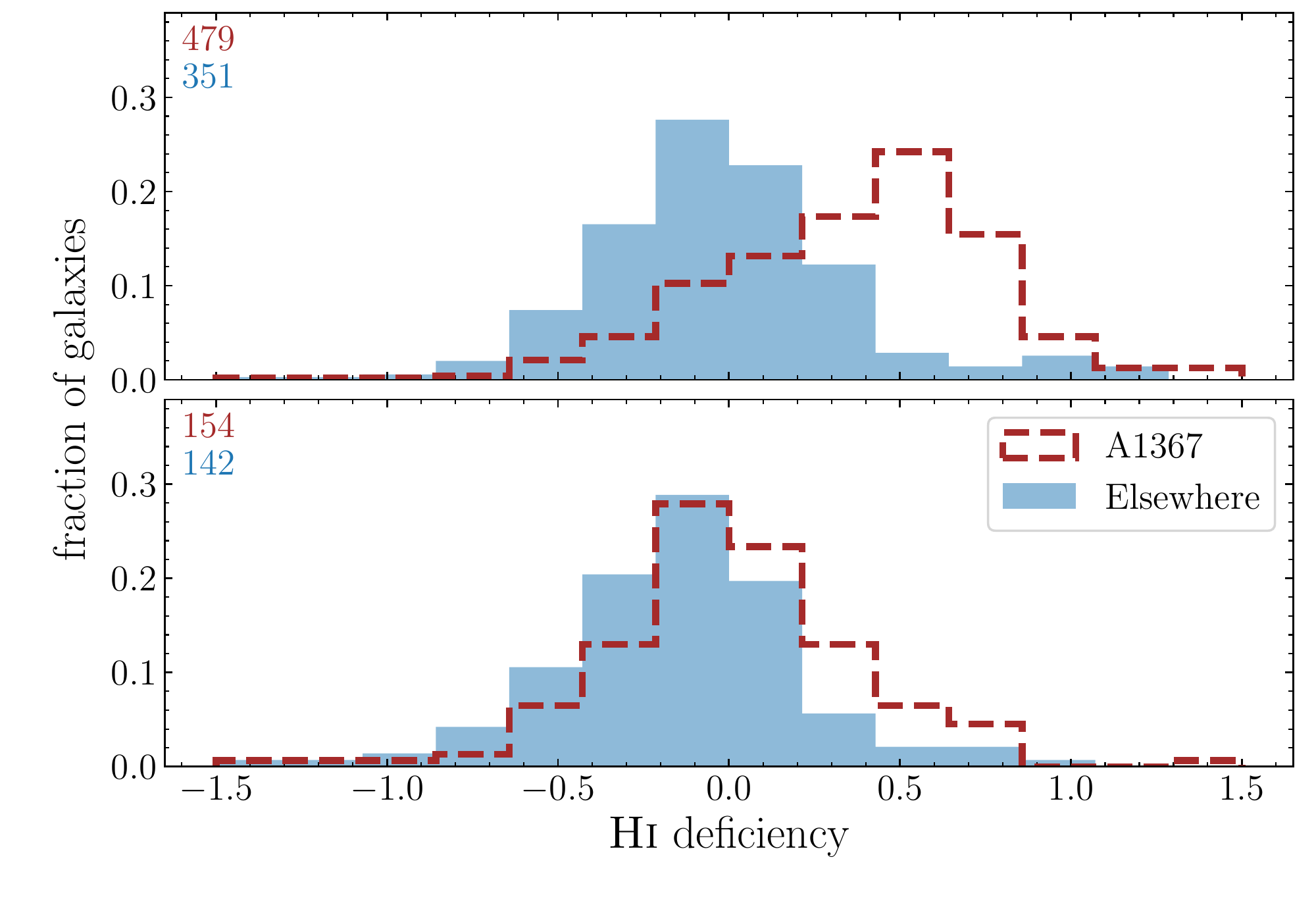}
      \caption{Distribution of \hi deficiency among all optical galaxies within the surveyed volume (top), and the \hi detections (bottom). Those are split into members of A1367 (brown, dashed histogram) and all the rest (blue, filled histogram). The total number of galaxies in each histogram is printed in the corresponding colour in the top left corner.\href{https://github.com/deshev/AGES-XI/blob/main/def_distr.py}{\includegraphics[height=3mm]{github-logo.png}}}
         \label{deficiency_distribution}
   \end{figure}
%-----------------------------------------------------------------

As defined by \citet{Haynes&Giovanelli1984} and used by many researchers \citep[e.g.][]{Cortese2011, Stark2016, Catinella2018, Loni2021}, the calculation of \hi deficiency relies on additional data and a calibration of the relation between optical diameter and \Mhi. This calibration is based on measurements of the optical diameter in a particular filter and limiting magnitude, and is known to depend on galaxy morphology. It also relies on a sample of galaxies with 'normal' gas content that is unaltered by external factors and observed through a telescope capable of measuring the emission present at all the spatial frequencies. Obviously these requirements and additional data are not going to be available for all \hi data sets and might prove problematic in the future when we measure the \hi content of galaxies at higher redshifts. It is possible that the local calibration of the mass -- size relation does not hold for high-redshift galaxies. All of the above-mentioned issues and the specifics of our deficiency estimate drive a relatively high uncertainty. For galaxies with DEF < 0.5, that is, those that within the uncertainty have the expected \hi content, we find a mean DEF = -0.005$\pm$ 0.33. A comparison with \citet{Cortese2011} shows good agreement with their estimate of 0.06$\pm$0.28.

The top panel of Fig.\ref{deficiency_distribution} presents the distribution of \hi deficiency for all galaxies with optical redshift measurements within the \hi survey. Those are split in two groups: members of A1367 and all others (which we term 'Elsewhere'), which have means of 0.41$\pm$0.49 and -0.02$\pm$0.41, respectively. A comparison between the \hi deficiencies of the cluster and the Elsewhere sample show that DEF = 0.3 is a good dividing point between normal and deficient galaxies, in agreement with \citet{Dressler1986}, \citet{Solanes2001}, \citet{Loni2021}. This means that 87\% of the Elsewhere sample can be considered \hi normal, as opposed to 38\% for the cluster sample.

It is obvious that the most \hi-deficient galaxies would not be detected at all at 21cm. The bottom panel of Fig.\ref{deficiency_distribution} shows the distribution of \hi deficiency only for \hi-detected galaxies. Removing the non-detections from the sample drastically reduces the difference between the means of the two environmental bins, from 0.43 to 0.12. The standard method for accounting for non-detected galaxies is to compute an upper limit on their detectable gas content to find the lower limit on their deficiency, as we did in Section \ref{section:deficiency_results} and present in Fig. \ref{mhimstar} and in the upper panel of Fig.\ref{deficiency_distribution}. This offers the advantage of giving information on individual objects, but relies on the calibration of the expected \hi mass from external data sets. We note that the $\sim$70\% reduction in the difference between the means of the two distributions can be accounted for by simply counting the non-detected fraction. The use of the non-detected fraction may offer a simpler alternative to the deficiency limits, and we explore the advantages and disadvantages of this parameter below.

\subsection{\hi 'non-detected fraction' and \hi deficiency}\label{section:det_fr_def}
%----------------------------------------------------------------- 
   \begin{figure*}
   \centering
   \includegraphics[width=\hsize]{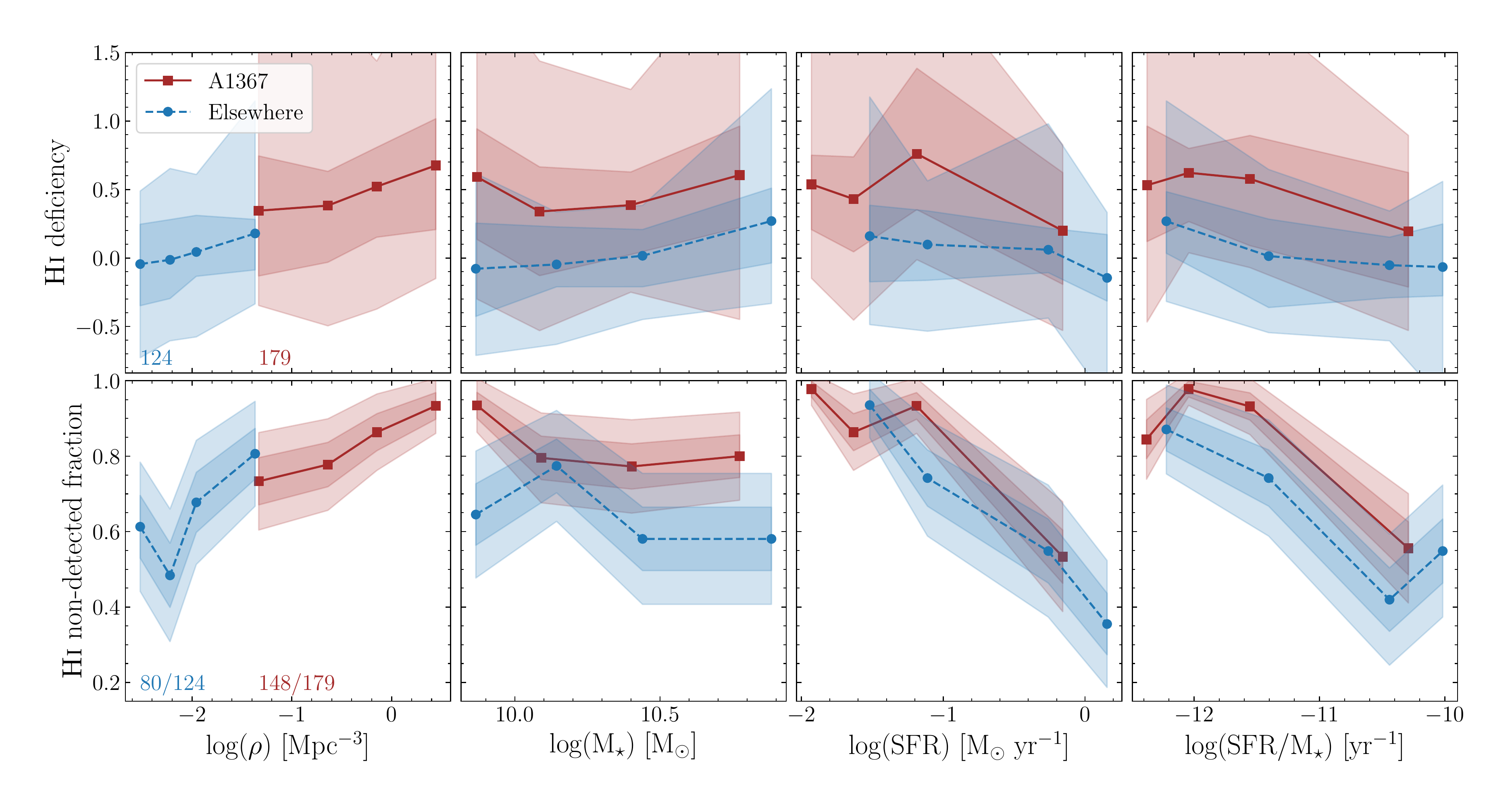}
      \caption{\hi deficiency (top) and \hi 'non-detected fraction' (bottom), as a function of (left to right) local galaxy density, stellar mass, star formation rate, and specific star formation rate. Cluster galaxies are shown with brown squares and solid lines, galaxies outside the cluster are shown with blue dots and dashed lines. The number of galaxies in the two environments is printed in the top left panel in the corresponding colour. The overall number of "non-detected" / all galaxies is printed in the bottom left panel. The shaded regions show 1$\sigma$ and 2$\sigma$ assuming normal distribution.\href{https://github.com/deshev/AGES-XI/blob/main/compare_HIDF_HIdef.py}{\includegraphics[height=3mm]{github-logo.png}}}
         \label{HIDF}
   \end{figure*}
%----------------------------------------------------------------- 

Because any observations will have a sensitivity limit varying with redshift, the 'non-detected fraction' needs to be calculated for a complete sample. In Fig.\ref{mass_distributions} we outline the stellar and \hi mass completeness limits as log \Mhi$^{lim}$ = 9.25 M$_{\odot}$ and log M$_{\star}^{lim}$ = 9.75 M$_{\odot}$. In this, section we analyse the 'non-detected fraction' as calculated for this stellar and \hi mass complete sample. In this sense, 'non-detected fraction' should be understood as an alias, meaning the fraction of galaxies with stellar mass above 9.75 and \hi mass below 9.25. We use the alias for brevity; it will be precisely equal to the true non-detected fraction for observations with uniform sensitivity. As Fig.\ref{mass_distributions} shows, many of the galaxies considered as 'non-detections' are in fact detected but have \Mhi below the completeness limit. As described earlier, these completeness limits apply to the AGES and SDSS data for sources with recession velocity $cz \leq$ 15500 km s$^{-1}$. Beyond this redshift, RFI pollution affects the completeness of the \hi survey.

The top panel of Fig.\ref{HIDF} shows the distribution of \hi deficiency for the stellar-mass-complete sample. The bottom row shows the \hi 'non-detected fraction' for the \Mhi and M$_{\star}$ complete sample. The number of galaxies in the two samples is printed in the first panels. These are presented as a function of local galaxy density, stellar mass, star formation rate and specific star formation rate. The shaded areas on the top row show the 2.5, 17, 83, and 97.5 percentile of the distribution of \hi deficiency (1$\sigma$ and 2$\sigma$ for a Gaussian distribution). The uncertainties shown on the bottom row are binomial 66\% and 95\% confidence intervals calculated using normal approximation. The points are binned along the x-axis in bins of constant population number to guarantee uniform significance along the relations. Along the x-axis the points are plotted at the median value for all galaxies in the bin. We choose this over the mean because of its reduced susceptibility to outliers, even though it makes the distribution appear narrower. 

As already apparent from Fig.\ref{polar_plot} and the bottom panel of Fig\ref{mass_distributions}, the cluster galaxies are distributed at higher local densities than the rest. There is, in fact, a small overlap between the two distributions of local densities which is hidden by the binning in the left-most panels of Fig.\ref{HIDF}. The Elsewhere sample contains a number of groups as well as void-like regions. The highest density region outside A1367 present in our data is at the high-redshift end of the survey which is heavily polluted by RFI from the radar at the nearby Luis Mu\~{n}oz Mar\'{i}n International Airport (see Fig. \ref{polar_plot}). As discussed above, this part is excluded from this analysis. We would like to point out that while this particular problem is caused by ground-based facilities, these types of problems for radio astronomy, in general, are set to become much more severe in the near future with the introduction of constellation satellites.

The left-most panels show that, qualitatively, \hi deficiency and the 'non-detected fraction' show the same trends. Both show a continuous increase with local density, irrespective of the global environment, that is, whether or not the galaxy belongs to a cluster. The \hi 'non-detected fraction' outside the cluster matches very well the average non-detected fraction published by \citet{Huang2012} based on the 40\% ALFALFA catalogue. As clusters of galaxies are very rare objects in the Universe and represent only a tiny fraction of the volume surveyed by ALFALFA, this is not surprising. \citet{Denes2014} published maps of \hi deficiency as a function of local galaxy density, showing essentially the same result as that shown in the left-most column of Fig.\ref{HIDF}, despite the different method of measuring local galaxy density.

The second column of Fig.\ref{HIDF} again shows a qualitative agreement between \hi deficiency and \hi 'non-detected fraction'. At a given stellar mass, cluster galaxies are more gas deficient than galaxies outside the cluster. This result is also in good qualitative agreement with the results by \citep{Cortese2011} and the results for satellites only by \citet{Stark2016}. Our sample does not allow segregation of satellites and centrals but we can assume that at least the first three mass bins are dominated by satellites and this defines the observed relations.

There is little difference between the third and fourth columns of Fig.\ref{HIDF}, which is expected given the lack of correlation with stellar mass. However, these panels appear to show a qualitative difference between \hi deficiency and the \hi 'non-detected fraction'. While the former shows only dependence on global environment but not on SFR, the latter shows no significant dependence on global environment but a strong correlation with SFR. However, it should be noted that because of the large scatter in the distribution of \hi deficiency it could also be considered consistent with the trend shown by the 'non-detected fraction'.

As we have already stressed, it is imperative that this analysis be based on a stellar and \hi mass complete sample. We verified that this is the case and also verified that the exact value at which completeness is achieved by the survey does not change the slope of the relation but only its intercept.

The 'non-detected fraction' is essentially a ratio between the integrals of the \hi mass function and stellar mass function above a certain limit. The exact limits used do not change the shape of the relation, only its normalisation, at least within this survey. Studies of the variations in the \hi mass function with environment point at a small increase in the knee of the Schechter function fit as the local density increases \citep{Jones2016} or of the slope of the low-mass end of the fit \citep{Zwaan2005}. Studies of the variations in the stellar mass function with environment are similarly inconclusive but generally agree that the high-mass slope increases in high-density environments while the low-mass slope likely decreases \citep{Etherington2017, Papovich2018}. Establishing the physics driving this relation requires larger data sets and will be the subject of a future paper. However, these results show the importance of mechanisms related to groups and clusters, like tidal interactions, mergers and/or ram pressure stripping. The local density, in this case, is measured as the inverse of the volume available for each galaxy. This is a parameter which could be related to the gas accretion rate from the circumgalactic medium. Availability of data for groups and clusters with a range of properties will be needed to establish whether it is active gas stripping or lack of gas accretion that shapes this relation.

The 'non-detected fraction' is in effect a low-resolution \hi deficiency which results in significantly reduced uncertainties compared to \hi deficiency. In the case of the correlation between SFR and the 'non-detected fraction' (Fig. \ref{HIDF}), it is obvious that the large scatter in \hi deficiency prevents us from seeing its correlation with SFR. Trading resolution for sensitivity is common practice when low S/Ns are involved; see for example Section \ref{section:stacking}. The main problem with using the 'non-detected fraction' as a parameter for studying environmental effects is that it can only be defined for an ensemble of galaxies, while \hi deficiency can be calculated for individual galaxies. As such it is applicable to large blind surveys like AGES, while deficiency is superior for targetted, high-S/N, observations.

\section{Summary}
We present the complete Arecibo Galaxy Environment Survey (AGES) survey of Abell 1367 and its surroundings. This is one of the most sensitive blind \hi surveys performed to date. We detect 457 \hi sources over the imaged 44000 Mpc$^{3}$. Within the SDSS and NED, we found 335 optical counterparts with spectroscopic redshift. Searching through the NED database, we found another 106 potential optical counterparts with unknown optical redshift. The presented \hi catalogue contains 16 sources without optical counterparts.

These \hi sources are distributed over a wide range of local densities and are therefore well suited for studying the effects of environment on galaxy evolution. Analysis of the data with respect to local and global environment reveals that the dependence of \hi properties on global environment is part of a continuous dependence on local density.

\begin{acknowledgements}
This work was supported by the Czech Science Foundation grant 19-18647S and the institutional project RVO 67985815. This research made use of Astropy,\footnote{\url{http://www.astropy.org}} a community-developed core Python package for Astronomy \citep{astropy:2013, astropy:2018}. This research made use of NASA's Astrophysics Data System Bibliographic Services. This research has made use of the NASA/IPAC Extragalactic Database (NED), which is funded by the National Aeronautics and Space Administration and operated by the California Institute of Technology. This research made use of APLpy, an open-source plotting package for Python \citep{aplpy2012, aplpy2019}.

This publication uses data generated via the Zooniverse.org platform, development of which is funded by generous support, including a Global Impact Award from Google, and by a grant from the Alfred P. Sloan Foundation.
\end{acknowledgements}

\bibliographystyle{aa} % style aa.bst
\bibliography{/home/tazio/works/references} % your references Yourfile.bib

%%%%%%%%%%%%%%%%%%%%%%%%%%%%%%%%%%%%%%%%%%%%%%%%%%%%%%%%%
% Appendix HI atlas
\begin{appendix}
\section{Atlas of \hi detections}\label{atlas}
In this Appendix, we show five example \hi detections spanning a range of galaxy properties and S/Ns. The complete atlas showing all \hi detections presented in this article can be downloaded from \url{https://www.dropbox.com/s/xffshfom5h0m3f0/AGES_A1367_atlas.tar.gz?dl=0}.\\

Description of the individual panels. Top row, from left to right:\\
1 - Moment 0 map. The velocity integral is highlighted in red in the bottom panel. The cyan circle has a diameter equal to the FWHM of Arecibo's beam and is centred at the catalogue coordinates of the source.\\

2 - Renzogram. For every channel, the contour is drawn at a significance level indicated at the top. Contours are colour coded according to their velocity as indicated in the colour bar. The contours are overlaid on a RGB image made from r, g, and i band images from the SDSS. The cyan circle is the same as in the previous panel. The image has the same sky coverage as panel 1.\\

3 - The same optical image used in panel 2. The blue circles show SDSS spectroscopic extragalactic targets with the numbers indicating their recession velocity. Orange pentagons indicate other \hi detections in the field. Their catalogue name and recession velocity are indicated. The cyan circle is as in panel 1. The panel has the same extent on the sky as the previous panels.\\

4 - The optical image shown in panels 2 and 3 but showing only the beam area indicated by the circle on the previous panels.\\

5 - A white light optical image showing the same part of the sky as panel 4. This was produced by integrating the three channels and with 'sqrt' stretch to boost the visibility of low-surface-brightness features. Different symbols and colours indicate optical counterparts. The cyan square indicates the primary optical counterpart (as listed in the online table). Magenta triangles show the secondary optical counterpart. Yellow hexagons show potential optical counterparts. The recession velocity is indicated, when available.\\

Bottom row:\\
\hi spectrum of the source. The thin blue line shows the spectrum used for measurements. This is extracted from Hanning 3 smoothed cube. The thick blue line shows a spectrum extracted from a Hanning 5 smoothed cube used when searching for detections. The thin grey lines show reference spectra extracted at eight locations two beams away from the centre of the source. The solid vertical line shows the velocity centre of the source. The two dashed-dotted lines show the W$_{50}$ width of the source. The two dotted lines show the W$_{20}$ of the source. On the left, a number of measured parameters of the source are written, many of which are also given in the online tables. From top to bottom: total \hi mass, total flux, S/N of the peak, central velocity, and redshift of the source, flag of the \hi source, asymmetry of the line profile, S/N of the source.

\begin{landscape}
\begin{figure}
\resizebox{24.5cm}{!}
{\centering\includegraphics{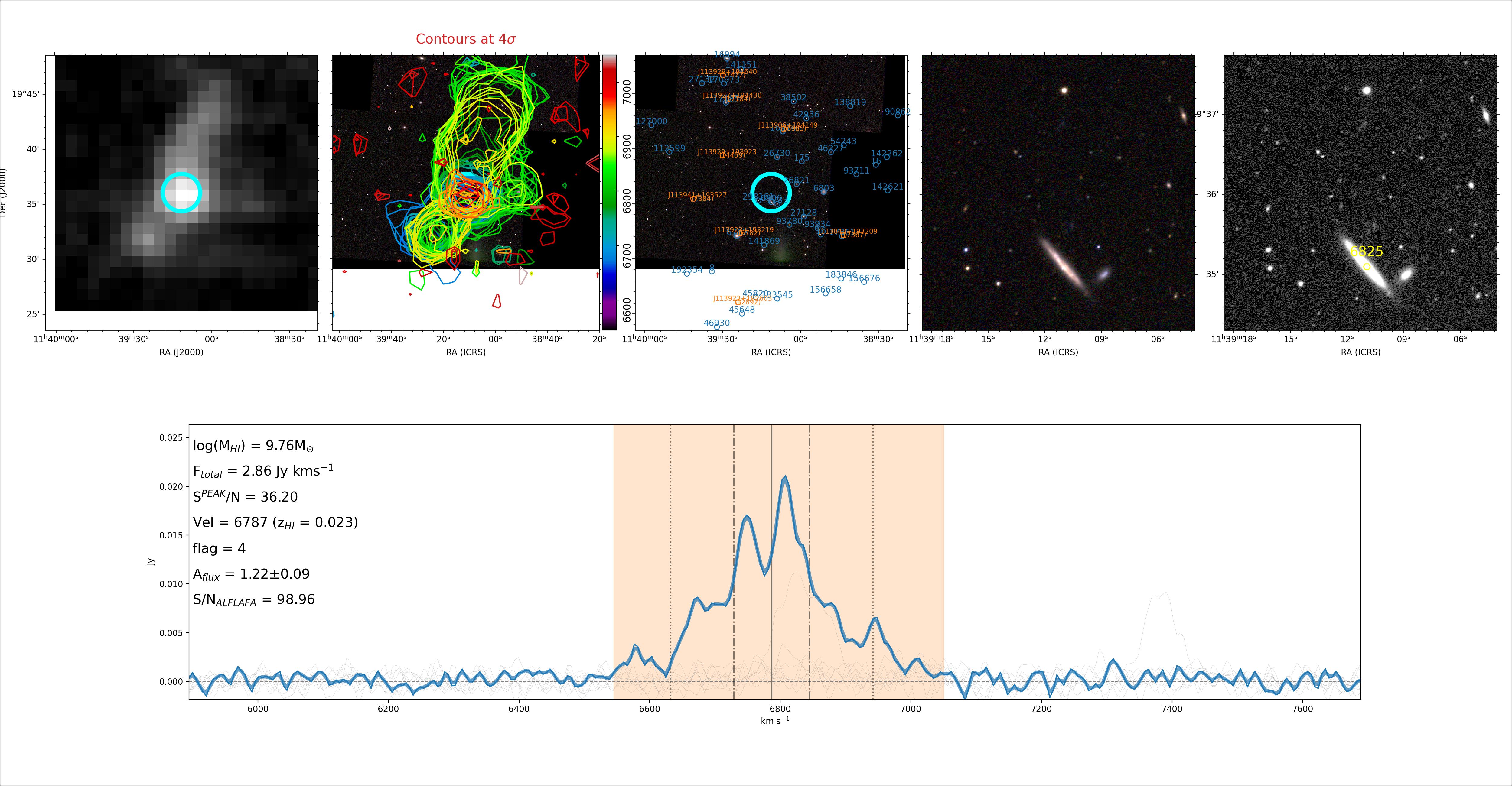}}
\caption{J113911+193601}
\end{figure}
\end{landscape}
\newpage

\begin{landscape}
\begin{figure}%f1
\resizebox{24.5cm}{!}
{\centering\includegraphics{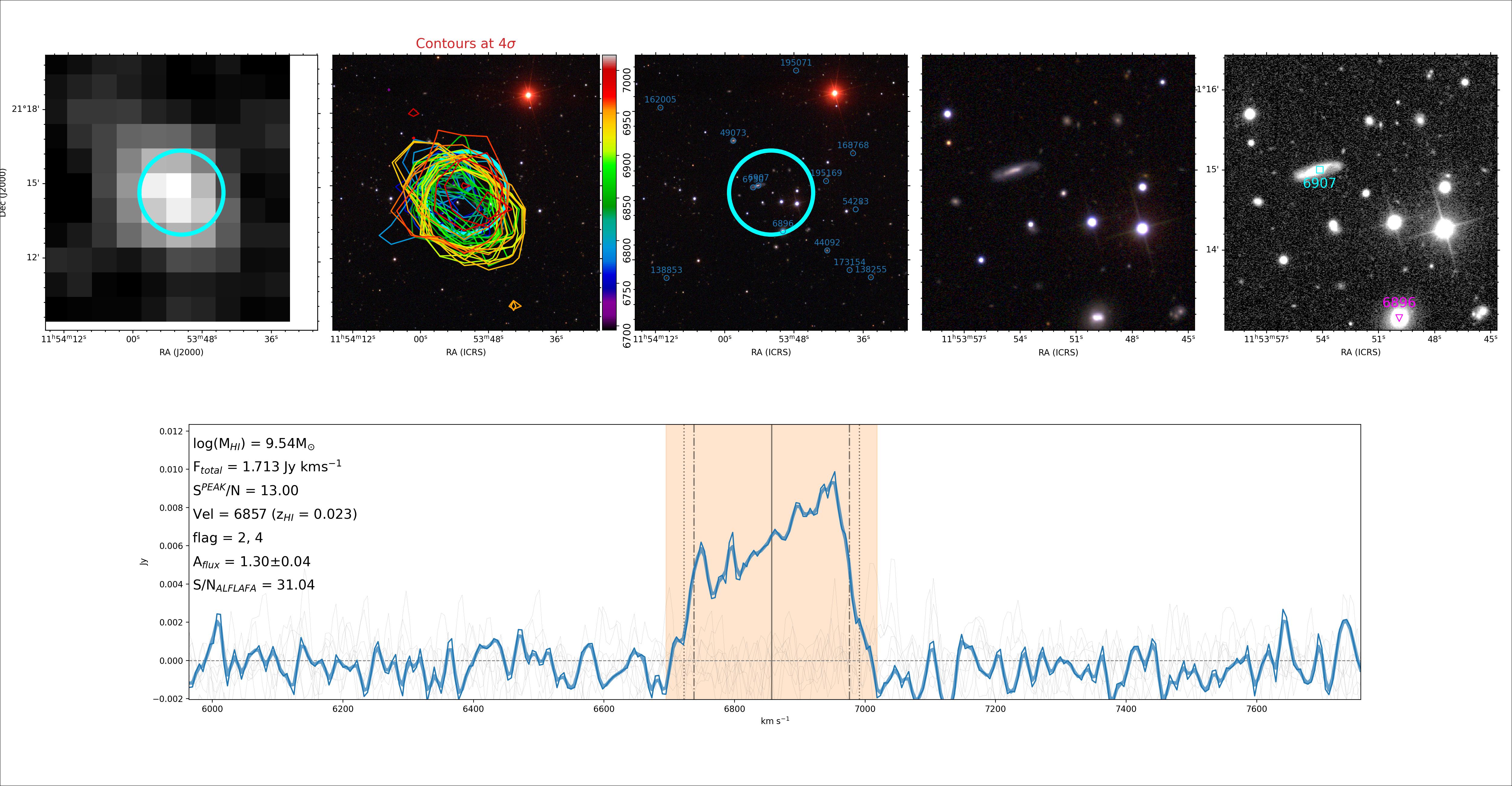}}
\caption{J115351+211443}
\end{figure}
\end{landscape}
\newpage

\begin{landscape}
\begin{figure}%[htp]%f1
\resizebox{24.5cm}{!}
{\centering\includegraphics{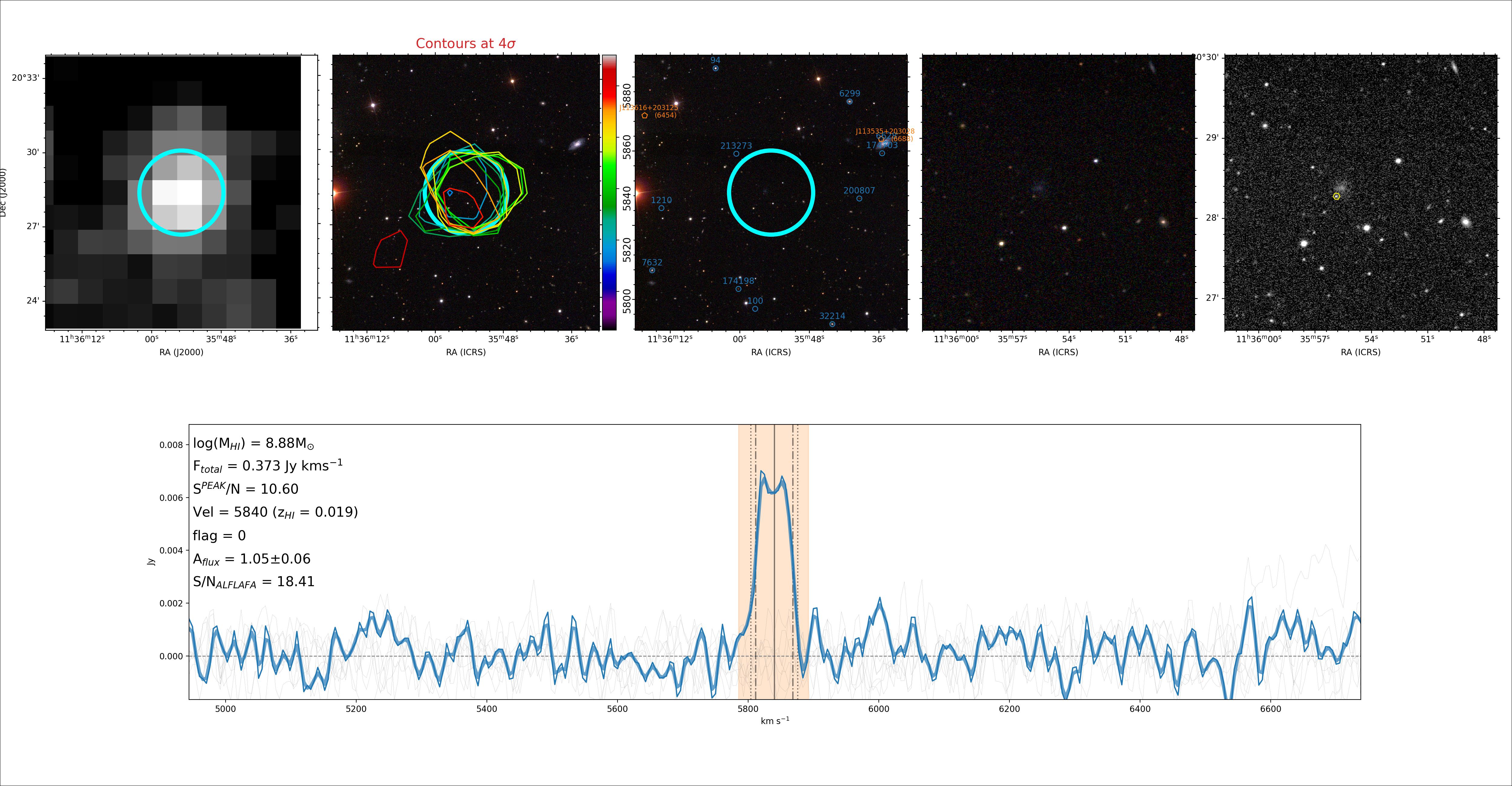}}
\caption{J113554+202819}
\end{figure}
\end{landscape}
\newpage

\begin{landscape}
\begin{figure}%f1
\resizebox{24.5cm}{!}
{\centering\includegraphics{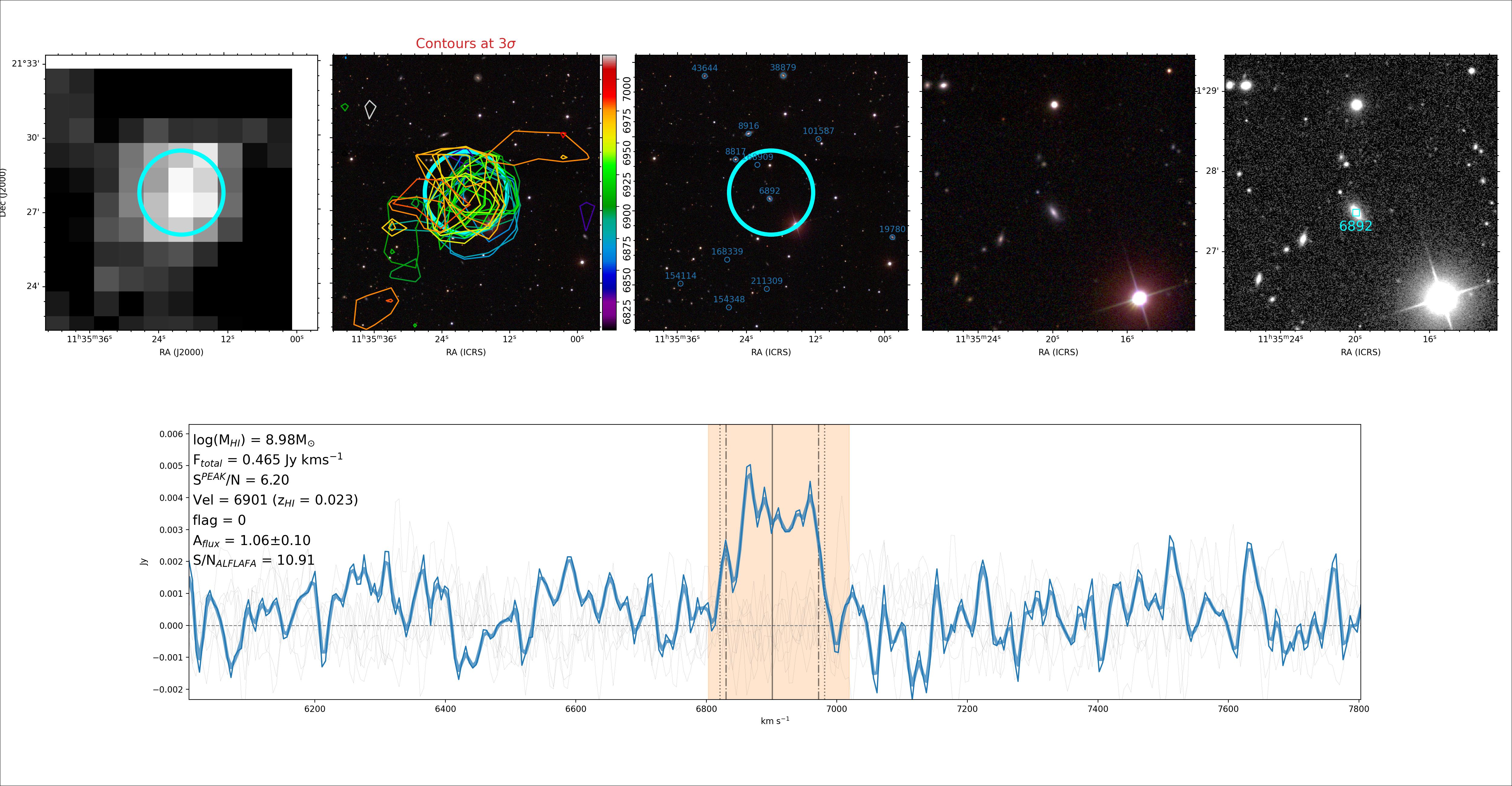}}
\caption{J113519+212744}
\end{figure}
\end{landscape}
\newpage

\begin{landscape}
\begin{figure}%f1
\resizebox{24.5cm}{!}
{\centering\includegraphics{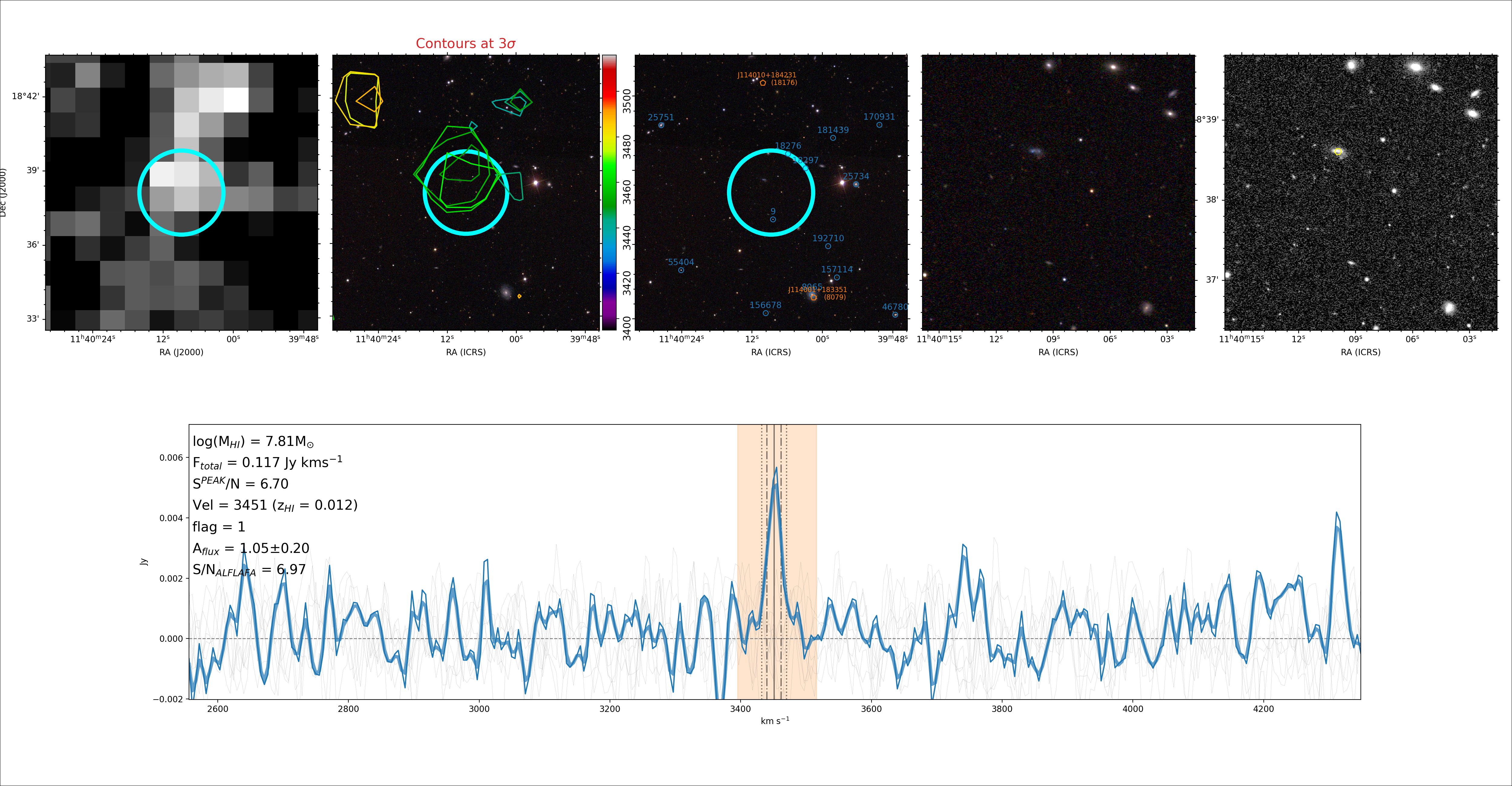}}
\caption{J114008+183805}
\end{figure}
\end{landscape}
\end{appendix}

\end{document}